\documentclass[aps,floatfix,reprint,prl,superscriptaddress]{revtex4-2}
\usepackage{graphicx,amsmath,amssymb,bbm,xcolor,soul}

\graphicspath{{main_text_figures/}{figures/}}

\begin{document}
\title{Topological Directional Coupler}

\author{Yandong Li}
 \email{yl2695@cornell.edu}
 \affiliation{School of Applied and Engineering Physics, Cornell University, Ithaca, New York, 14853, USA}
\author{Minwoo Jung}
 \affiliation{Department of Physics, Cornell University, Ithaca, New York, 14853, USA}
\author{Yang Yu}
 \affiliation{School of Applied and Engineering Physics, Cornell University, Ithaca, New York, 14853, USA}
\author{Yuchen Han}
 \affiliation{Department of Physics, Cornell University, Ithaca, New York, 14853, USA}
\author{Baile Zhang}
 \affiliation{Division of Physics and Applied Physics, School of Physical and Mathematical Sciences, Nanyang Technological University, 21 Nanyang Link, Singapore, 637371, Singapore}
 \affiliation{Centre for Disruptive Photonic Technologies, The Photonics Institute, Nanyang Technological University, 50 Nanyang Avenue, Singapore, 639798, Singapore}
\author{Gennady Shvets}
 \email{gshvets@cornell.edu}
 \affiliation{School of Applied and Engineering Physics, Cornell University, Ithaca, New York, 14853, USA}

\date{\today}

\begin{abstract}
Interferometers and beam splitters are fundamental building blocks for photonic neuromorphic and quantum computing machinery.
In waveguide-based photonic integrated circuits, beam-splitting is achieved with directional couplers that rely on transition regions where the waveguides are adiabatically bent to suppress back-reflection.
We present a novel, compact approach to introducing guided mode coupling.
By leveraging multimodal domain walls between microwave topological photonic crystals, we use the photonic-spin-conservation to suppress back-reflection while relaxing the topological protection of the valley degree of freedom to implement tunable beam splitting. Rapid advancements in chip-scale topological photonics suggest that the proposed simultaneous utilization of multiple topological degrees of freedom could benefit the development of novel photonic computing platforms.
\end{abstract}

\maketitle
Neuromorphic~\cite{YShen:2017} and quantum~\cite{NHarris:2017,Xanadu:2021} computing procedures involve a variety of linear algebra operations on a set of $N$-dimensional unitary matrices [U$(N)$].
Such operations can be carried out using basic photonic devices --- phase shifters and beam splitters.
Specifically, an arbitrary $N$ dimensional linear operation can be realized with $N(N-1)/2$ Mach-Zehnder interferometers (MZIs)~\cite{Zeilinger:1994,Milburn:2007,Clements:2016}.
Although it is challenging to perfectly align all the photonic elements with a free-space optical setup when $N$ becomes large, with guided-wave implementations enabled by modern integrated photonics techniques, all elements can be squeezed on a chip, thus motivating the development of large-scale MZI arrays~\cite{OBrien:2015,Waks:2020,Bogaerts:2020}.
However, key components of MZIs, directional couplers, sacrifice compactness to minimize the photon loss because the coupled waveguides need long (tens of wavelengths) transition regions to prevent unwanted back-reflections~\cite{HongTang:2016}.

Topological photonics offers new strategies for light manipulation, particularly suppressing back-reflections~\cite{Lu:2014,Shvets:2017,RMP:2019}: the immediate result of topological protection is granted by the conservation of the relevant topological index.
For example, first-order $\mathbb{Z}$ and $\mathbb{Z}_2$ indices (e.g., the Chern, spin-Chern, and valley-Chern numbers) guarantee robust propagation of edge states~\cite{Soljacic:2009,Khanikaev:2013,Nori:2015,Rechtsman:2018_1} at the domain walls between two topologically distinct bulks of photonic crystals (PhCs) with overlapping band gaps.
Applications of topological photonics have also been extended to maintaining the robustness of few-photon quantum states~\cite{Segev:2018,Blanco_Redondo:2019,Hafezi:2018,XFRen:2021}.

Inspired by the semiconductor pioneer H. Kroemer's famous phrase, ``the interface is the device"~\cite{Kroemer:2001}, we aim to design compact directional couplers by utilizing multimodal topological interfaces that support edge modes characterized by more than one topological index~\cite{Ezawa:2013}.
One candidate for the bulk PhC is a spin photonic crystal (SPC)~\cite{Khanikaev:2013,Ma:2017,YLi:2022}, whose band gaps are characterized by half-integer spin and valley Chern numbers $C_{s,v}=\pm 1/2$, where $s \in \{\uparrow,\downarrow\}$ and $v \in \{\mathbf{K}, \mathbf{K}'\}$ label the spin and valley degrees of freedom (DoFs), respectively.
A domain wall between two SPCs with opposite spin Chern numbers supports two chiral topological edge states in each valley.
Their propagation direction is locked to the spin DoF and protected from backscattering by the conservation of the spin DoF.
On the other hand, the valley DoF depends on the direction of the domain wall:
$\mathbf{K} (\mathbf{K}')$ valleys are well-defined at zigzag domain walls, and the two co-propagating edge states (one per valley) are protected against inter-valley scattering~\cite{Ma:2016,Rechtsman:2018_1,Gao:2018}.

When the protection of the valley DoF is intentionally broken in a controlled manner such that the spin DoF remains conserved, the two co-propagating states can crosstalk without experiencing backscattering.
Such intentional breaking of some (but not all) topological DoFs offers an opportunity for creating topological directional couplers (TDCs).
Harnessing the topological protection to suppress backscattering can potentially make TDCs smaller than their conventional counterparts because (a) the transition regions~\cite{HongTang:2016} between coupled waveguides become unnecessary;
(b) The strong modal overlap between the two valley-polarized states reduces the required length of the crosstalking region.
In other words, photonic spin conservation enables unidirectional propagation;
Valley hybridization causes a significant (non-evanescent) coupling between the two valley-polarized states and enables a short coupling length.
In this Letter, we demonstrate that by arranging the domain wall along the armchair direction and engineering its properties, inter-valley scattering is induced~\cite{Dresselhaus:1996,Fertig:2006,Rechtsman:2018_1,YLi:2020} in a controllable way, which enables a compact, tunable TDC.

\begin{figure}
\centering
    \includegraphics[width=0.48\textwidth]{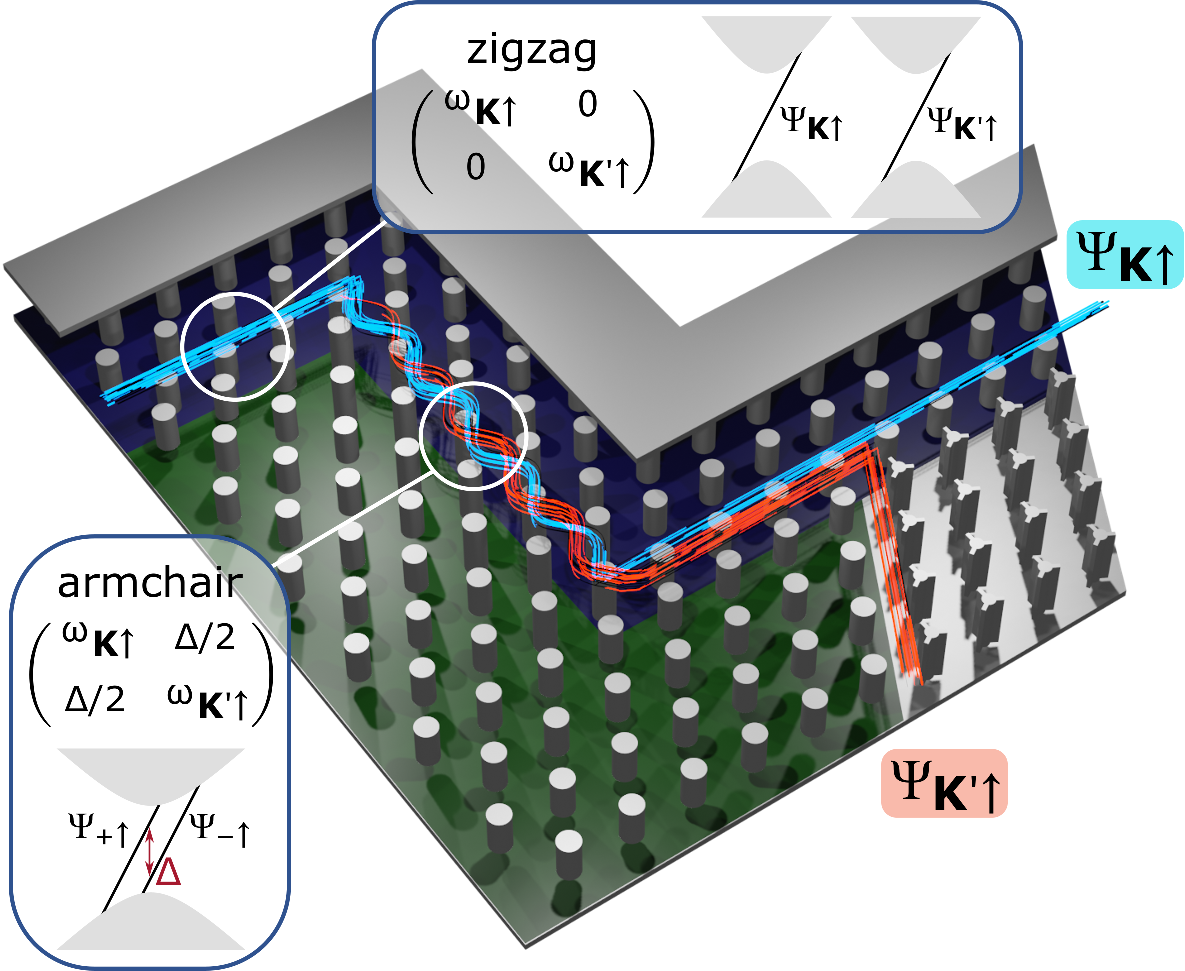}
    \caption{Topological directional coupler based on a designer domain wall between SPCs.
    Green (blue) regions: SPC$^{1(2)}$, metallic rods are attached to the top (bottom) metal plate.
    The input state $\Psi_{\mathbf{K} \uparrow}$ is injected in a zigzag domain wall, where the valley DoF is conserved.
    Along the armchair domain wall, the valley basis $\Psi_{\mathbf{K}(\mathbf{K}') \uparrow}$ hybridizes into $\Psi_{+(-) \uparrow}$.
    The bottom right gray region: VPC for routing valley-polarized modes for detection.
    Insets: Along the zigzag domain wall, $\mathbf{K}$ and $\mathbf{K}'$ valleys are separated in the $k$-space, and inter-valley scattering is negligible.
    The edge-state Hamiltonian is hence diagonal;
    Along the armchair domain wall, $\mathbf{K}$ and $\mathbf{K}'$ valleys are folded to the $\Gamma$ point, resulting in a strong inter-valley scattering, and the off-diagonal components in the edge-state Hamiltonian become nonzero.
    }
    \label{fig:1}
\end{figure}

We use the microwave SPCs introduced in Ref.~\cite{Ma:2017,Gao:2018} to implement such a TDC.
As depicted in Fig.~\ref{fig:1}, the structure is based on the domain wall between SPC$^{1(2)}$, triangular lattices of metallic rods attached to the top (bottom) radiation-confining metallic plates.
The photonic spin DoF is synthesized by mixing the TE- and TM-polarized modes, which is accomplished by breaking the reflection symmetry about the mid-height $x-y$ plane~\cite{Ma:2017}.
The SPC$^{1,2}$ possess spin-Chern numbers $C_s^{(1,2)} = \mp 1/2$, accordingly (where the superscript $(1)$ corresponds to $-1/2$ for spin up and $1/2$ for spin down).
Details of the physical setup are described in the Supplementary Material.
We judiciously arrange the SPC$^{1,2}$ and let them form a domain wall with a $90^{\circ}$ turning corner -- the domain wall is zigzag-type before, and armchair-type after the turn (Fig.~\ref{fig:1}).
Along zigzag domain walls, the discrete translational symmetry along one of the primitive lattice vector directions ($\mathbf{a}_1$ or $\mathbf{a}_2$) is conserved, and therefore the valley DoF is conserved.
In contrast, armchair domain walls are along the $\mathbf{a}_1 + \mathbf{a}_2$ direction, thus changing the discrete translational symmetry and folding both valleys to the $\Gamma$ point ($k=0$).

For the zigzag domain wall, the edge-state Hamiltonian is diagonal in the valley basis, $\left( \Psi_{\mathbf{K} \uparrow}, \Psi_{\mathbf{K}' \uparrow} \right)^T$, because the two valleys are well-separated in the $k$-space.
When an arbitrary spin-up mode $\Psi_{\uparrow}= \cos(\theta/2) \Psi_{\mathbf{K} \uparrow} + e^{i\varphi} \sin(\theta/2) \Psi_{\mathbf{K}' \uparrow}$ is guided into or excited in the zigzag domain wall,
the proportion of the $\mathbf{K}(\mathbf{K}')$ valley-polarization,
which is the modulus squared projection onto the valley-basis, $P^{\text{zig}}_{\mathbf{K}(\mathbf{K}') \uparrow}(l) \equiv |\langle \Psi^{\text{zig}}_{\uparrow} (l) | \Psi_{\mathbf{K}(\mathbf{K}') \uparrow} \rangle|^2$, is independent of the propagation distance $l$,
\begin{equation}\label{eq:zigzag}
\begin{aligned}
& P^{\text{zig}}_{\mathbf{K} \uparrow}(l) = \cos^2 \theta/2, \\
& P^{\text{zig}}_{\mathbf{K}' \uparrow}(l) = \sin^2 \theta/2.
\end{aligned}
\end{equation}
Here, only the spin-up modes are considered because the spin-down ones exit the structure after excitation.

On the other hand, for the armchair domain wall, the edge state Hamiltonian can be represented as $H_{\text{arm}} = \left( \omega_0, \Delta/2; \Delta/2, \omega_0 \right)$, where $\omega_0 \equiv \omega_{\mathbf{K} \uparrow} = \omega_{\mathbf{K}' \uparrow}$, because the two Dirac cones at the $\mathbf{K}$ and $\mathbf{K}'$ valleys are gapped through the same mechanism.
The off-diagonal term $\Delta/2$ represents the $\mathbf{K}$-$\mathbf{K}'$ inter-valley scattering that hybridizes the valley-polarized modes:
The guided mode basis changes from $\left( \Psi_{\mathbf{K} \uparrow}, \Psi_{\mathbf{K}' \uparrow} \right)^T$ to $\left( \Psi_{+ \uparrow}, \Psi_{- \uparrow} \right)^T$, where $\Psi_{\pm \uparrow} \equiv \left( \Psi_{\mathbf{K} \uparrow} \pm \Psi_{\mathbf{K}' \uparrow} \right)/\sqrt{2}$.
When $\Psi_{\uparrow}$ enters the armchair domain wall, its time-evolution is governed by the operator $e^{-i H_{\text{arm}} l/v_g}$.
Hence, $\Psi_{\uparrow}$ experiences a Rabi oscillation, and the proportion of the $\mathbf{K}(\mathbf{K}')$ valley polarization, $P^{\text{arm}}_{\mathbf{K}(\mathbf{K}') \uparrow}(l) \equiv |\langle \Psi^{\text{arm}}_{\uparrow} (l) | \Psi_{\mathbf{K}(\mathbf{K}') \uparrow} \rangle|^2$, becomes
\begin{equation}\label{eq:armchair}
\begin{aligned}
& P^{\text{arm}}_{\mathbf{K} \uparrow}(l) = \cos^2 \left( \theta/2 \right) \cos^2 \left[ \Phi(\Delta) \right] + \sin^2 \left( \theta/2 \right) \sin^2 \left[ \Phi(\Delta) \right], \\
& P^{\text{arm}}_{\mathbf{K}' \uparrow}(l) = \cos^2 \left( \theta/2 \right) \sin^2 \left[ \Phi(\Delta) \right] + \sin^2 \left( \theta/2 \right) \cos^2 \left[ \Phi(\Delta) \right],
\end{aligned}
\end{equation}
where $\Phi(\Delta) \equiv l\Delta/\left( 2v_g \right)$.
Hence, the armchair domain wall rotates $\Psi_{\uparrow}$ in the valley basis.

In order to measure the two proportions $P_{\mathbf{K}(\mathbf{K}') \uparrow}(L)$ at the end of the armchair domain wall of length $l=L$, we first restore the valley-DoF-conservation.
We let the domain wall make another $90^{\circ}$ turn, and the armchair domain wall out-couples to a zigzag one, along which valley-conservation ensures that $P_{\mathbf{K}(\mathbf{K}') \uparrow}(L)$ remains unchanged.
Then, we joint a valley photonic crystal (VPC) with the two SPCs to form two single-mode topological waveguides (Fig.~\ref{fig:1}), and each waveguide supports only one valley-polarization for the spin-up mode~\cite{Ma:2017,Kang:2018}.
Therefore, the two valley-polarizations are separated, and $P_{\mathbf{K}(\mathbf{K}') \uparrow}(L)$ can be calculated by measuring the output energy of the two single-mode waveguides~\cite{YLi:2022}.

To summarize our design, the engineered domain wall performs two changes-of-basis.
The guided edge mode basis changes from $\left( \Psi_{\mathbf{K} \uparrow}, \Psi_{\mathbf{K}' \uparrow} \right)^T$ to $\left( \Psi_{+ \uparrow}, \Psi_{- \uparrow} \right)^T$ and back to $\left( \Psi_{\mathbf{K} \uparrow}, \Psi_{\mathbf{K}' \uparrow} \right)^T$ again, corresponding to the three different types of domain walls: zigzag SPC$^1$-SPC$^2$, armchair SPC$^1$-SPC$^2$, and zigzag SPC$^{1,2}$-VPC.
This design realizes a unitary transformation, $e^{-i H_{\text{arm}} L/v_g}$, on valley-polarized topological edge modes, $\Psi_{\mathbf{K}(\mathbf{K}') \uparrow}$.
The form of this unitary transformation offers two tuning knobs: one is $L$, the length of the armchair domain wall, which, however, requires significant changes of the structure dimension.
The other is the degree of inter-valley scattering, which is represented by $\Delta$ in the off-diagonal entries of $H_{\text{arm}}$ and can be tuned by varying unit cell parameters in the vicinity of the domain wall.

\begin{figure}
\centering
    \includegraphics[width=0.49\textwidth]{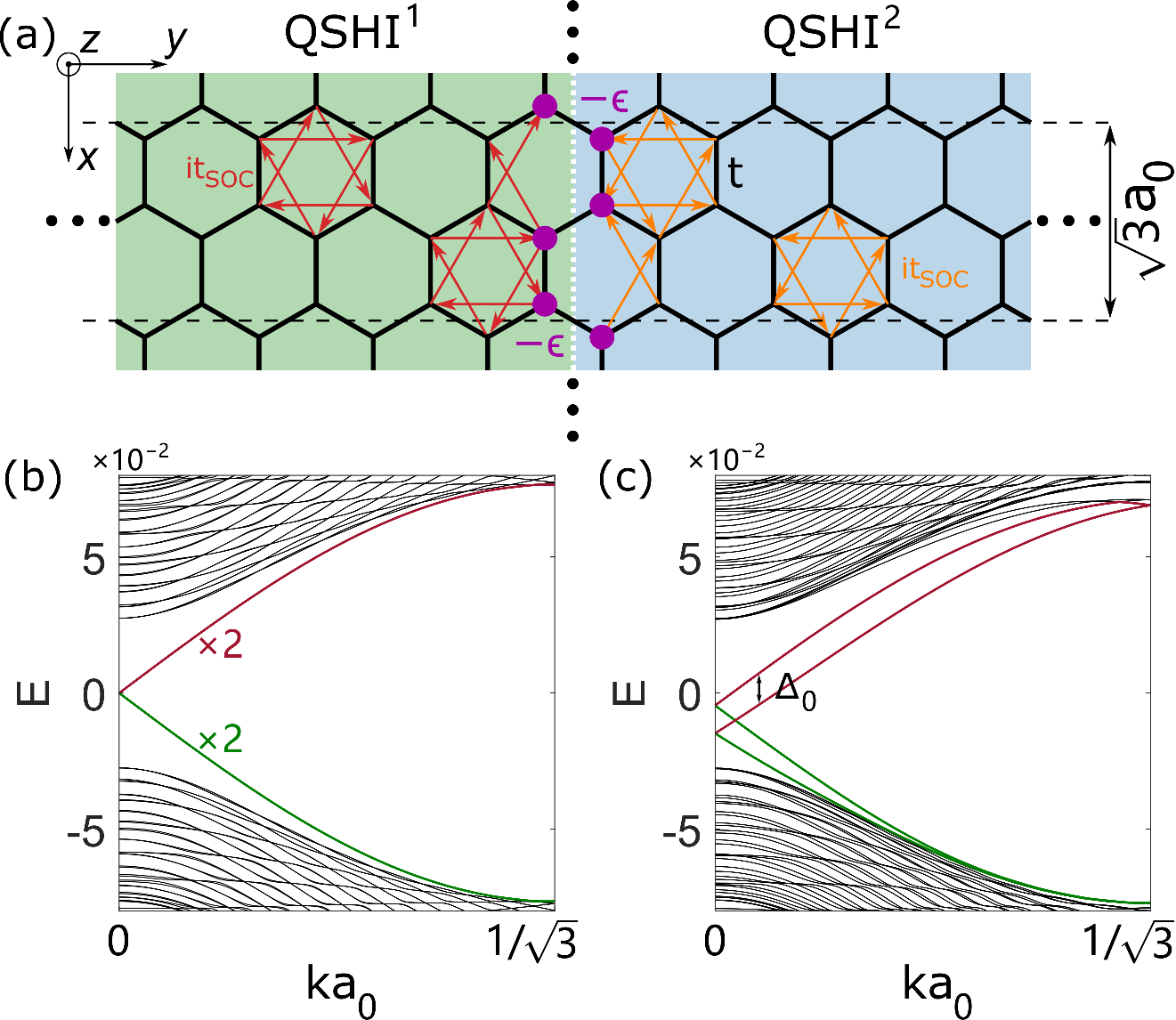}
    \caption{The tight-binding model.
    (a) Schematic of the armchair domain wall between two 2D quantum spin Hall insulators (QSHIs).
    In the green (blue) region, we implement clockwise (counterclockwise) next nearest neighbor (NNN) hoppings [red (orange) arrows] for spin up.
    For spin down, the NNN hoppings are reversed.
    The atomic sites adjacent to the domain wall are perturbed with the on-site potential $-\epsilon$ (purple dots).
    (b,c) Band diagrams of the armchair domain wall with the perturbation (b) $\epsilon = 0$ and (c) $|\epsilon| = 0.42|t|$.
    In (b), the spin up (down) modes are doubly degenerate.
    $|t|$ is the nearest neighbor hopping strength.
    }
    \label{fig:2}
\end{figure}

The inter-valley scattering amplitude is sensitive to detailed changes along the armchair domain wall.
Here, we illustrate the tuning of the $\mathbf{K}$-$\mathbf{K}'$-valley hybridization using a tight-binding model.
In this model, the quantum spin Hall and quantum valley Hall effects are implemented on a honeycomb lattice with the Kane-Mele model~\cite{KaneMele:2005} and a staggered on-site potential~\cite{Semenoff:1984}, respectively.
All parameters are determined according to the band structure of the corresponding PhCs (see the Supplementary Materials for details).
In this idealized model, there is no inter-valley scattering along the unperturbed ($\epsilon = 0$) armchair domain wall, and the spin up (down) edge modes are doubly degenerate.
To introduce inter-valley scattering, we perturb the armchair domain wall by applying the on-site potential $-\epsilon$ on the atomic sites along the armchair domain wall.
This perturbation hybridizes the valley-polarized modes $\Psi_{\mathbf{K}(\mathbf{K}') \uparrow}$ to $\Psi_{\pm \uparrow}$ (Fig.~\ref{fig:2}).
The energy-splitting between the two hybridized modes, $\Delta \equiv \omega_{+ \uparrow} - \omega_{- \uparrow}$, increases linearly with $|\epsilon|$.
Specifically, when $|\epsilon| \approx 0.42 |t|$, $\Delta = \Delta_0$, where $\Delta_0 \approx 1.1 \times 10^{-2} (2\pi c/a_0)$ (this perturbation is chosen to be consistent with the inter-valley scattering obtained in our experiment when $g=g_0$, Fig.~\ref{fig:4}) and $t$ is the nearest-neighbor hopping.

\begin{figure}
\centering
    \includegraphics[width=0.49\textwidth]{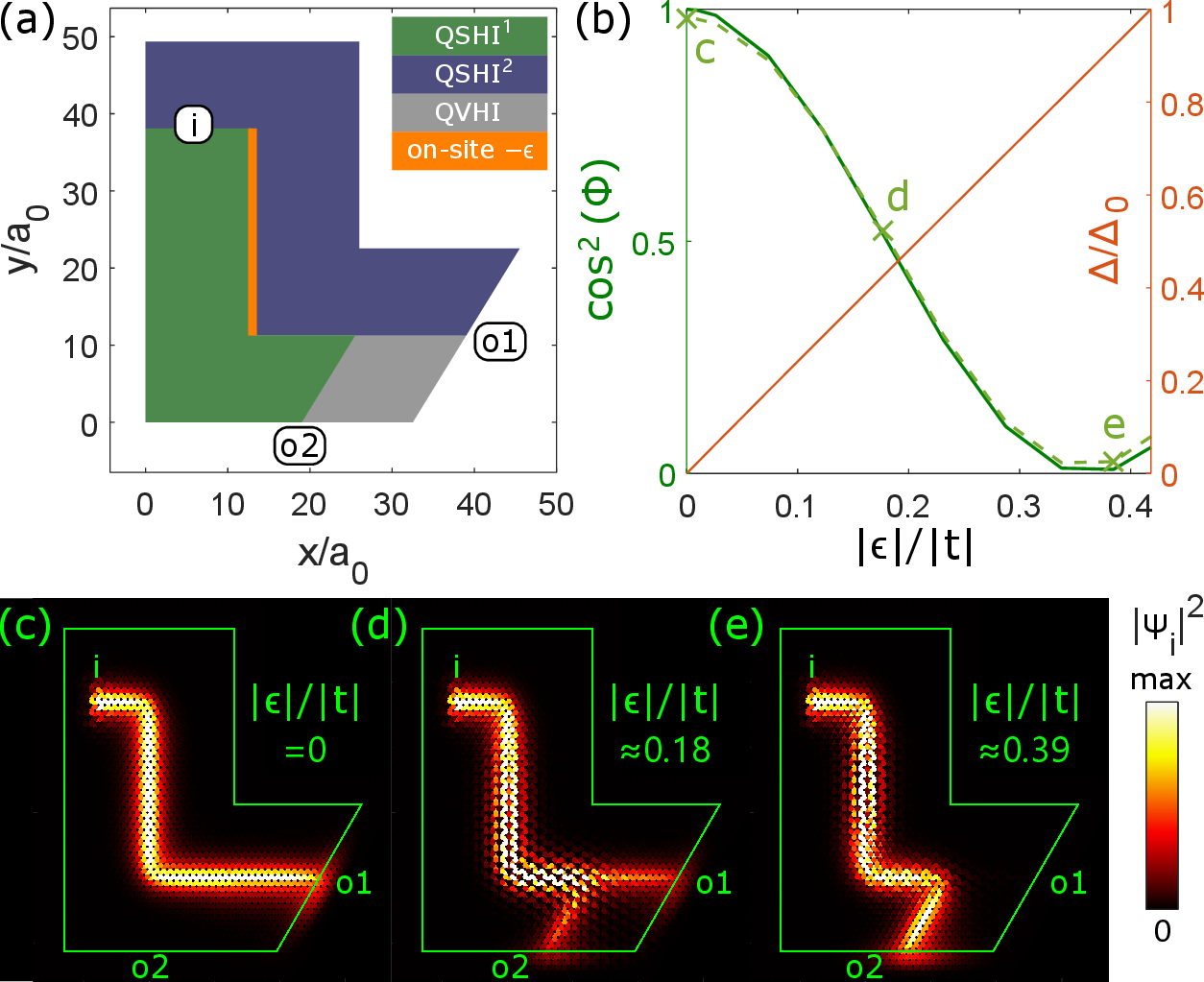}
    \caption{Tight-binding simulation of the tunable Rabi oscillation.
    (a) Schematic of the tight-binding model.
    The green (blue) region represents QSHIs$^{1(2)}$ with clockwise (counter-clockwise) next nearest neighbor hopping.
    The gray region represents the QVHI.
    The orange line along the armchair domain wall represents the atomic sites that experience the perturbation $-\epsilon$.
    (b) Analytical (solid line) and tight-binding (dashed line) results of tuning the $\mathbf{K}$-valley proportion in the output by varying the band-splitting $\Delta$.
    $\Delta_0$ corresponds to the inter-valley scattering with $|\epsilon| \approx 0.42|t|$ along the armchair domain wall.
    (c-e) Field profiles of Rabi oscillations with different inter-valley scattering $\Delta$.
    After exiting the coupling region (armchair domain wall), the state becomes (c) $\mathbf{K}$-valley polarized, $P^{\text{TB}}_{\mathbf{K} \uparrow} : P^{\text{TB}}_{\mathbf{K}' \uparrow} \approx 1 : 0$;
    (d) equally $\mathbf{K}$- and $\mathbf{K}'$-valley polarized, $P^{\text{TB}}_{\mathbf{K} \uparrow} : P^{\text{TB}}_{\mathbf{K}' \uparrow} \approx 0.5 : 0.5$;
    (e) $\mathbf{K}'$-valley polarized, $P^{\text{TB}}_{\mathbf{K} \uparrow} : P^{\text{TB}}_{\mathbf{K}' \uparrow} \approx 0 : 1$.
    }
    \label{fig:3}
\end{figure}

The tunable $\mathbf{K}$-$\mathbf{K}'$-valley coupling at the armchair domain wall manifests itself when we measure the proportions $P_{\mathbf{K} \uparrow}$ and $P_{\mathbf{K}' \uparrow}$ after the Rabi oscillation.
In the tight-binding simulation, we judiciously implemented a source that excites the $\Psi_{\mathbf{K} \uparrow}$ state only.
$P_{\mathbf{K} \uparrow}$ and $P_{\mathbf{K}' \uparrow}$ are calculated using the squared amplitude $|\psi|^2$ at the output ports of the two single-mode valley-polarized waveguides, o1 and o2, according to the valley polarization that each waveguide selects:
$P^{\text{TB}}_{\mathbf{K}(\mathbf{K}') \uparrow} \equiv \sum_{i \in \text{o1(o2)}}|\psi_i|^2 / \sum_{i \in \{\text{o1,o2}\}}|\psi_i|^2$.
When $|\epsilon| = 0$, inter-valley scattering is completely suppressed ($\Delta=0$), and the valley DoF is conserved.
In this case, almost all the energy is directed to output port o1 and $P^{\text{TB}}_{\mathbf{K} \uparrow} : P^{\text{TB}}_{\mathbf{K}' \uparrow} \approx 1 : 0$, confirming that the output state is $\mathbf{K}$-valley polarized (Fig.~\ref{fig:3}c).
When $|\epsilon| \approx 0.18 |t|$, $\Phi(\Delta) \approx \pi/4$ in Eq.~\ref{eq:armchair},
and the output state has equal valley proportions, $P^{\text{TB}}_{\mathbf{K} \uparrow} : P^{\text{TB}}_{\mathbf{K}' \uparrow} \approx 0.5 : 0.5$ (Fig.~\ref{fig:3}d).
Finally, when $|\epsilon| \approx 0.39 |t|$, the inter-valley scattering at the armchair domain wall becomes large enough to result in $\Phi(\Delta) \approx \pi/2$, completely flipping the valley polarization from $\mathbf{K}$ to $\mathbf{K}'$.
Consequently, the output is $\mathbf{K}'$-valley polarized, $P^{\text{TB}}_{\mathbf{K} \uparrow} : P^{\text{TB}}_{\mathbf{K}' \uparrow} \approx 0 : 1$ (Fig.~\ref{fig:3}e).
The tight-binding model hence confirms that one type of perturbation along the armchair domain wall tunes the inter-valley scattering, and the setup in Fig.~\ref{fig:3}a can be viewed as a lattice model of a tunable TDC.

\begin{figure}
\centering
    \includegraphics[width=0.48\textwidth]{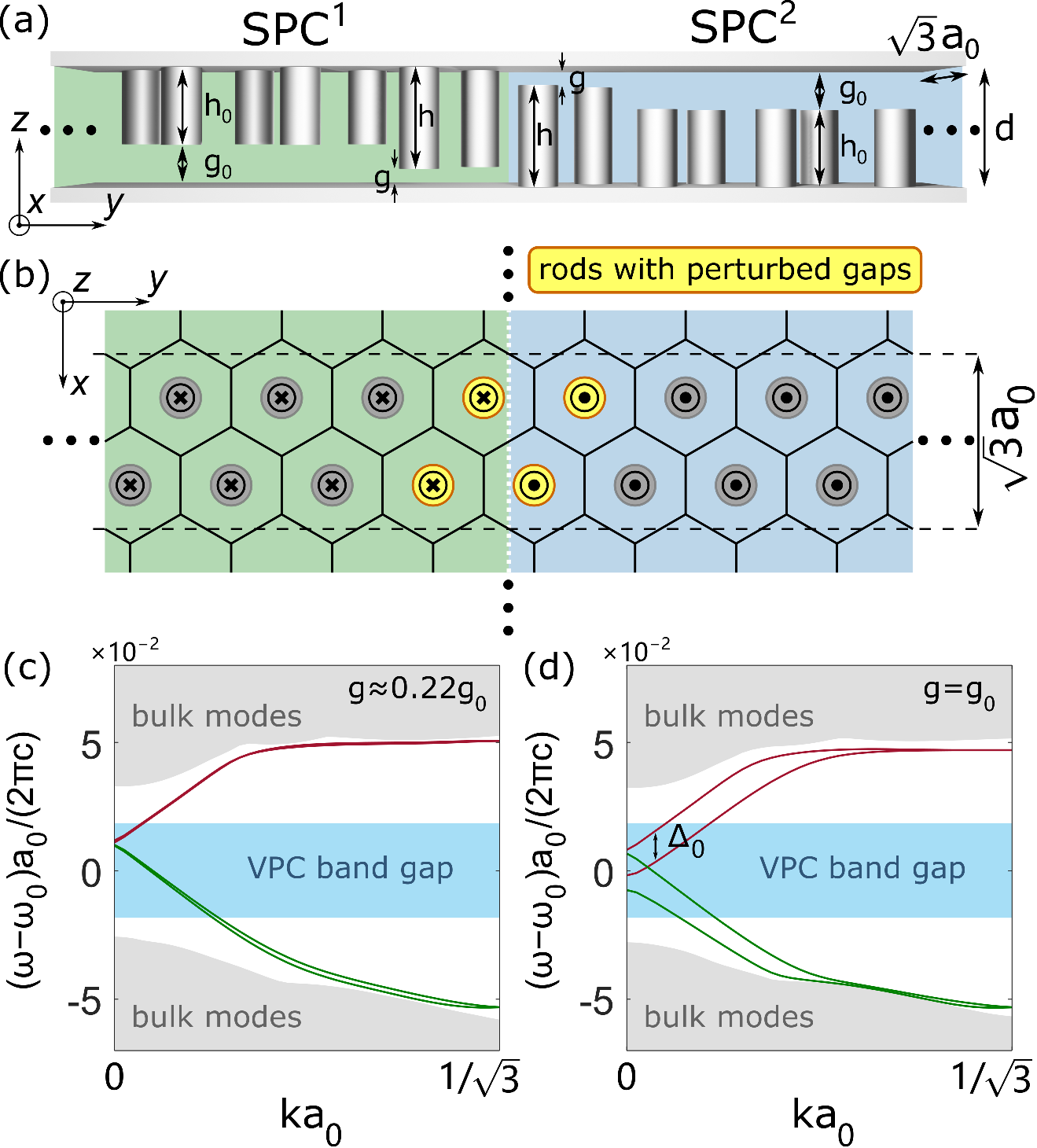}
    \caption{(a,b) The photonic implementation of the armchair domain wall between QSHI$^{1,2}$ using SPC$^{1,2}$.
    The gaps $g$ adjacent to the domain wall are perturbed.
    (c,d) Photonic edge modes supported by the armchair domain wall with (c) $g \approx 0.22g_0$ and (d) $g = g_0$.
    Parameters of SPC$^{1,2}$: $g_0 = 0.15a_0$ is the height of unperturbed gaps;
    $h_0 = 0.85a_0$, $d=a_0$, where $a_0$ is the lattice constant.
    The mid-gap frequency of the VPC band gap is $\omega_0 \approx 0.75 (2\pi c/a_0)$.
    }
    \label{fig:4}
\end{figure}

To introduce tunability in the photonic implementation of the TDC (Fig.~\ref{fig:1}), we use the height of the gaps between the rods and the top (bottom) plate and most adjacent to the domain wall, $g$ (Fig.~\ref{fig:4}a), as the control knob for manipulating the $\mathbf{K}$-$\mathbf{K}'$-valley scattering because the electromagnetic field of edge modes is exponentially localized at the domain wall and predominantly concentrated in those gaps.
By extending the cavity perturbation theory to incorporate multimodal interactions~\cite{Slater:1946,Dombrowski:1984}, we find that the effects of perturbing those gaps are mostly confined to each spin subspace -- the spin DoF is largely conserved when the valleys are hybridized (see the Supplementary Material for details).
When $g \approx 0.22 g_0$, the inter-valley scattering is negligible, corresponding to $\Delta \approx 0$ (Fig.~\ref{fig:4}c).
As $g$ increases, the inter-valley scattering is enhanced.
When $g=g_0$, the inter-valley scattering at the armchair domain wall causes a strong $\mathbf{K}$-$\mathbf{K}'$-valley hybridization with the splitting $\Delta = \Delta_0$ (Fig.~\ref{fig:4}d)
(Note that $\Delta_0$ shares the same value obtained from the tight-binding model with the perturbation setting $|\epsilon| \approx 0.42 |t|$.)

\begin{figure}
\centering
    \includegraphics[width=0.48\textwidth]{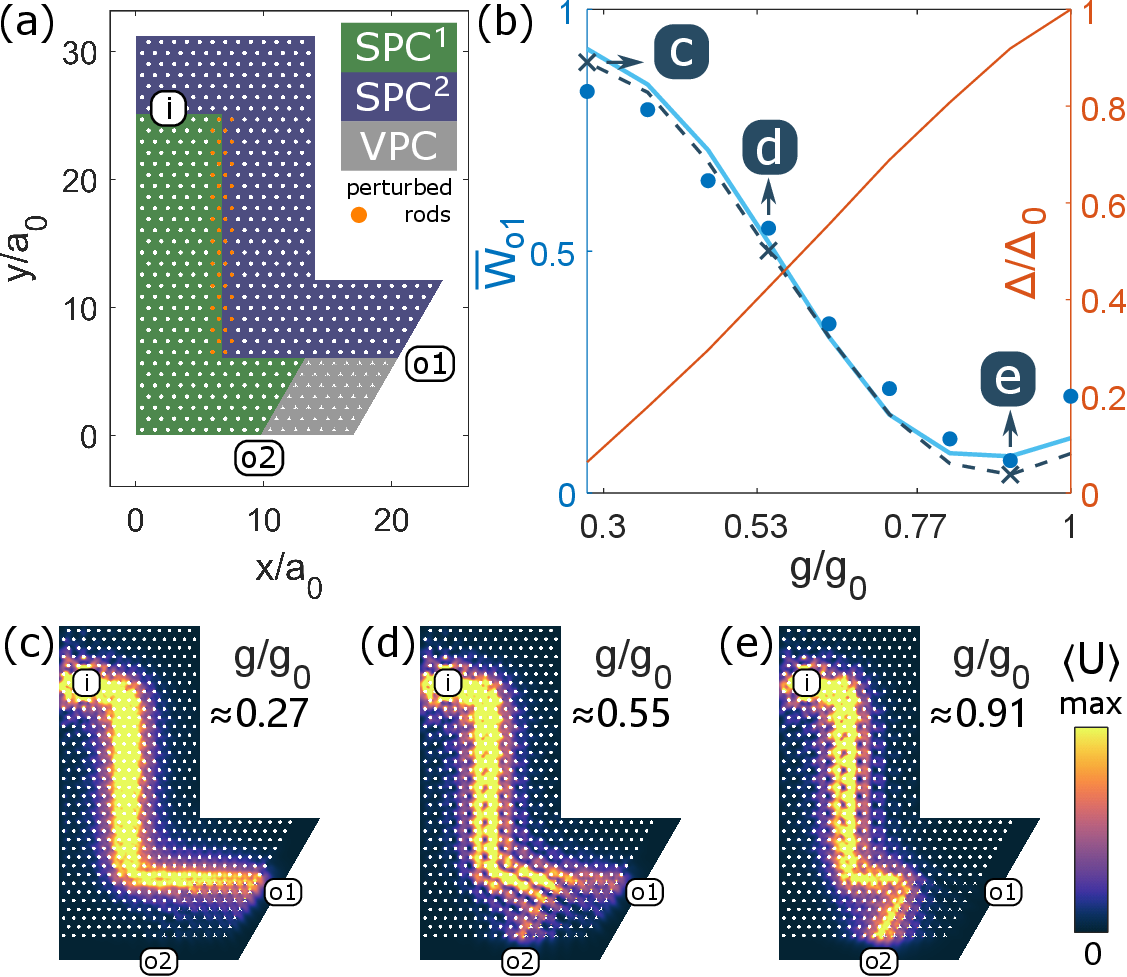}
    \caption{The experimental implementation and electromagnetic simulation of the TDC.
    (a) Schematic of the TDC based on topological PhCs.
    Blue (green) region: SPC$^2$ (SPC$^1$);
    Gray region: VPC.
    Orange dots along the armchair domain wall: rods with the perturbed gaps $g$.
    (b) Semi-analytical (solid line), simulation (dashed line), and experiment (dots) results of the electromagnetic energy ratio received at output port o1.
    $\overline{W}_{\text{o1}}$ is associated with the $\mathbf{K}$-valley polarization.
    $\Delta_0$ corresponds to the inter-valley scattering with $g = g_0$ in Fig.~\ref{fig:4}a.
    The experimental data is averaged over the frequency range $\omega_0 < \omega < 1.022\omega_0$, where back-reflection is suppressed (see Supplementary Material for details).
    (c-e) COMSOL simulation of the beam splitting with three different $g$ values at $\omega \approx 1.01\omega_0$:
    (c) $W_\text{o1}:W_\text{o2} \approx 0.89:0.11$;
    (d) $W_\text{o1}:W_\text{o2} \approx 0.50:0.50$;
    (e) $W_\text{o1}:W_\text{o2} \approx 0.04:0.96$.
    }
    \label{fig:5}
\end{figure}

The experimental implementation of the TDC was carried out in the microwave regime with $a0=36.8$mm (see Fig.~\ref{fig:4} for other parameters).
We increase the $\mathbf{K}$-$\mathbf{K}'$-valley hybridization by incrementing $g$ from $0.27 g_0$ to $g_0$.
We calculate the proportions of the valley polarizations, $P_{\mathbf{K} \uparrow}$ and $P_{\mathbf{K}' \uparrow}$, using the electromagnetic field energy at output ports o1 and o2: $P^{\text{exp}}_{\mathbf{K}(\mathbf{K}') \uparrow} = \overline{W}_{\text{o1(o2)}} \equiv W_{\text{o1(o2)}} / \sum_{i \in \{1,2\}} W_{\text{o}i}$.
Experimental results demonstrate that, as the gap height $g$ increases, $P^{\text{exp}}_{\mathbf{K} \uparrow} : P^{\text{exp}}_{\mathbf{K}' \uparrow}$ changes from $0.83:0.17$ to $0.07:0.93$, agreeing with COMSOL simulations (Fig.~\ref{fig:5}b-e).
In contrast to the tight-binding model, $P_{\mathbf{K} \uparrow} : P_{\mathbf{K}' \uparrow}$ of the photonic structure does not reach $1:0$ and $0:1$.
Two factors give rise to this phenomenon: the imperfectness of the source and the finite dimension of the structure.
The excited state is not exactly $\mathbf{K}$-valley-polarized, instead, the excitation efficiency ratio is $\eta_{\mathbf{K} \uparrow} / \eta_{\mathbf{K}' \uparrow} \approx 43$ for the realistic source.
On the other hand, due to the finite dimension of the VPC, the detection antenna at output port o1 not only receives the transmitted $\Psi_{\mathbf{K} \uparrow}$ but also the evanescent tail of the tunneled $\Psi_{\mathbf{K}' \uparrow}$.
Taking both factors into consideration, our semi-analytical model extending Eq.~\ref{eq:armchair} agrees with both simulation and experiment (see the Supplementary Material and Ref.~\cite{YLi:2022} for details).

We have demonstrated a tunable TDC based on tuning the valley-hybridization of edge states via relaxing the topological protection of the valley DoF.
With additional delay lines, any arbitrary U$(2)$ operation on the valley-polarized states can be realized using SPC$^{1,2}$ and VPC as elementary building components (see Supplementary Material for details).
When the valley-hybridization is tuned, the spin DoF stays conserved for a wide range of frequency (from $\omega_0$ to $1.022\omega_0$, about $50\%$ of the VPC band gap width), and the guided modes remain immune to back-scattering while turning around $90^{\circ}$ corners, keeping the structure compact.
Moreover, because the edge states supported by the multi-mode waveguiding domain wall originate from the Dirac cones gapped through the same mechanism, they propagate at identical group velocities across the topological band gap, potentially accommodating more complicated light flow manipulations.

Our design offers a new way for developing interferometers -- through breaking the protection of one specific topological DoF, the corresponding edge modes are coupled.
This paradigm offers a control knob from first principles for adjusting the coupling: the extent of topological-protection-breaking.
Therefore, our design is fundamentally different from conventional directional couplers, for which mode hybridization is due to the evanescent coupling of the two waveguide modes and usually tuned by empirically adjusting the waveguide geometry and separation~\cite{Crespi:2013}.
Additionally, TDCs can scale up with a cascading architecture, similar to MZI meshes.
Compared to interferometers based on single-mode topological waveguides~\cite{XFRen:2021}, our design harnesses the multi-modeness and offers controllability via carefully relaxing the valley conservation, both of which could enable applications in microwave metamaterials and quantum computing~\cite{TJCui:2022,Gu_Review:2017}.
With all-dielectric SPCs and VPCs for the telecommunication spectrum~\cite{Kang:2018,Litchinitser:2019}, the same design can possibly be realized on chip and combined with other topological photonic components~\cite{LiangFeng:2023,Hafezi:2020,YLi:2020,Litchinitser:2019,QJWang:2020}.
Furthermore, with light-controlled index-programmable materials~\cite{LiangFeng:2019,TWu:2023,MoLi:2024,Onodera:2024}, the perturbation along the armchair channel could be tuned remotely, paving the way for future progress in integrated photonics, photonic neuromorphic computing, and programmable quantum circuits~\cite{Wetzstein:2020,Thompson:2020,NP_Review:2022}.

\section{References}
\bibliographystyle{unsrt}

\pagebreak
\widetext
\begin{center}
\textbf{\large Supplementary Material for\\ ``Topological Directional Coupler''}
\end{center}
%%%%%%%%%% Merge with supplemental materials %%%%%%%%%%
%%%%%%%%%% Prefix a "S" to all equations, figures, tables and reset the counter %%%%%%%%%%
\setcounter{equation}{0}
\setcounter{figure}{0}
\setcounter{table}{0}
\setcounter{page}{1}
\makeatletter
\renewcommand{\theequation}{S\arabic{equation}}
\renewcommand{\thefigure}{S\arabic{figure}}
%\renewcommand{\bibnumfmt}[1]{[S#1]}
%\renewcommand{\citenumfont}[1]{S#1}
%%%%%%%%%% Prefix a "S" to all equations, figures, tables and reset the counter %%%%%%%%%%

\section*{Derivation of the two-edge-state Hamiltonian}

The Kane-Mele Hamiltonian can be written as
\begin{equation}
\mathcal{H} = t \sum_{\langle m,n \rangle} a^{\dagger}_m b_n \sigma^0_{\alpha\beta} + it_{SOC} \sum_{\langle\langle m,n \rangle\rangle} \nu_{mn} (a^{\dagger}_{m \alpha} a_{n \beta} + b^{\dagger}_{m \alpha} b_{n \beta}) \sigma^z_{\alpha\beta},
\end{equation}
where $a^{\dagger}_j(b^{\dagger}_j)$ and $a_j(b_j)$ are the creation and annihilation operators for the mode at the A(B)-typed site with index $j$.
$\nu_{mn} = 1$ if the next nearest neighbor (NNN) coupling is clockwise about the lattice center;
$\nu_{mn} = -1$ if it is counterclockwise.
The spin labels, $\alpha,\beta \in \left\{ \uparrow, \downarrow \right\}$, and $\sigma^0_{\alpha\beta}, \sigma^z_{\alpha\beta}$ are Pauli matrices acting on the spin subspace.
$t, t_{SOC}$ are the nearest neighbor coupling and the NNN spin-orbit coupling, respectively~\cite{SM_KaneMele:2005}.

The corresponding $\mathbf{k}$-space Hamiltonian in the spin up subspace is
\begin{equation}\label{eq:H_k}
H_{\uparrow}(\mathbf{k}) =
\begin{pmatrix}
t_{SOC} g(\mathbf{k}) & t f(\mathbf{k}) \\
t f^*(\mathbf{k}) & -t_{SOC} g(\mathbf{k})
\end{pmatrix},
\end{equation}
where $f(\mathbf{k}) = e^{i k_y a_0/\sqrt{3}} + 2\cos(k_x a_0/2) e^{-i \sqrt{3} k_y a_0/6}$
and $g(\mathbf{k}) = 2[\sin (k_x a_0) - 2\sin (k_x a_0/2) \cos (\sqrt{3}k_y a_0/2)]$.
For convenience, we set $a_0=1$ in the following derivation.

In the vicinity of the valleys $\mathbf{K}=(-4\pi/3, 0)$ and $\mathbf{K}'=(4\pi/3, 0)$, Eq.~\ref{eq:H_k} can be expanded with respect to $\delta \mathbf{k} = (\delta k_x, \delta k_y)$ as,
\begin{equation}\label{eq:H_deltak}
H_{\mathbf{K}\uparrow} = v(\delta k_x \sigma_x + \delta k_y \sigma_y) + m \sigma_z
\quad \text{and} \quad
H_{\mathbf{K}'\uparrow} = v(-\delta k_x \sigma_x + \delta k_y \sigma_y) - m \sigma_z,
\end{equation}
where $v = \sqrt{3} t/2$ and $m = 3\sqrt{3} t_{SOC}$.

Eq.~\ref{eq:H_deltak} can be combined into a form similar to the BHZ Hamiltonian~\cite{SM_BHZ:2006},
\begin{equation}
H_{\uparrow} \equiv
\begin{pmatrix}
H_{\mathbf{K}\uparrow} & 0 \\
0 & V^{-1} H_{\mathbf{K}'\uparrow} V
\end{pmatrix}
= v (\delta k_x \tau_z \otimes \sigma_x + \delta k_y \tau_z \otimes \sigma_y) + m \tau_0 \otimes \sigma_z.
\end{equation}
The basis is $\left( \psi^R_{\mathbf{K} \uparrow}, \psi^L_{\mathbf{K} \uparrow}, \psi^R_{\mathbf{K}' \uparrow}, \psi^L_{\mathbf{K}' \uparrow} \right)^T$.
$R$ ($L$) represents the RCP (LCP) state.
$V = \left( 0, 1; 1, 0 \right)$ is the rotation matrix that exchanges the RCP and LCP bases.

We solve for the edge state $\Psi$ at the interface between two bulks with opposite-signed $t_{SOC}$~\cite{SM_Schindler:2020,SM_TIbook:2015,SM_YLi:2022},
\begin{equation}
\begin{aligned}\label{eq:EFA}
& [- \hat{p}_x \sigma_x + \hat{p}_y \sigma_y + m(y) \sigma_z] \Psi(x,y) = E\Psi(x,y), \\
& m(y)=
\begin{cases}
m, & \text{for}\ y>0, \\
-m, & \text{for}\ y<0,
\end{cases}
\end{aligned}
\end{equation}
where $\hat{p}_x = -i\partial / \partial x$, $\hat{p}_y = -i\partial / \partial y$.
The dispersion relations of two edge states corresponding to the $\mathbf{K}$ and $\mathbf{K}'$ valleys are linear functions with an identical, positive slope,
\begin{equation}\label{eq:edgestate}
\omega_{\mathbf{K} \uparrow} = v (k_x + 4\pi/3 + 2\pi N) \quad \text{and} \quad
\omega_{\mathbf{K}' \uparrow} = v (k_x - 4\pi/3 + 2\pi N),
\end{equation}
where $N \in \mathbb{Z}$, $k_x$ is the wavevector along the interface.

Overall, the entire topological-insulator-based system is reduced to an effective two-state system of the two guided spin-up valley-polarized modes, with which we demonstrate the controllable interference.
At the zigzag interface, the two edge states, $\Psi_{\mathbf{K} \uparrow}$ and $\Psi_{\mathbf{K}' \uparrow}$, are well-separated in $k$-space, so they interact negligibly.
On the other hand, at the armchair interface, the two edge states are both folded to be centered at $k_x=0$ and interact with each other through inter-valley scattering.
Consequently, they hybridize into $\Psi_{\pm \uparrow} = (\Psi_{\mathbf{K} \uparrow} \pm \Psi_{\mathbf{K}' \uparrow}) /\sqrt{2}$.
The two-state Hamiltonians describing the two valley-polarized modes are
\begin{equation}
H_{\text{zigzag}} =
\begin{pmatrix}
\omega_{\mathbf{K} \uparrow} & 0 \\
0 & \omega_{\mathbf{K'} \uparrow}
\end{pmatrix}
\quad \text{and} \quad
H_{\text{armchair}} =
\begin{pmatrix}
\omega_{\mathbf{K} \uparrow} & \Delta/2 \\
\Delta/2 & \omega_{\mathbf{K'} \uparrow}
\end{pmatrix}.
\end{equation}

\section*{Determining tight-binding parameters}

We determine the tight-binding parameters by matching the tight-binding and the normalized photonic band structures.
Based on table, $|t_{SOC}|/|t| \approx 6.5 \times 10^{-2}$, and $|t_0|/|t| \approx 2.6 \times 10^{-1}$.

\begin{table}[ht]
 \caption{calculated tight-binding model parameters}
 \begin{tabular}{|c || c | c|} 
 \hline
 physical quantity & value in photonic band diagram & tight-binding parameter \\ [0.5ex]
 \hline\hline
 slope of the ungapped Dirac cone, $\frac{a_0}{2\pi c} \frac{\partial \omega}{\partial k}$ & $6.6 \times 10^{-2}$ & $\sqrt{3}|t|/2$ \\ [0.5ex]
 \hline
 band gap width of bulk SPC at $\mathbf{K}$($\mathbf{K}'$), $\frac{a_0}{2\pi c} \Delta\omega_{\text{SPC}}$ & $5.2 \times 10^{-2}$ & $6\sqrt{3} |t|_{SOC}$ \\ [0.5ex]
 \hline
 band gap width of bulk VPC at $\mathbf{K}$($\mathbf{K}'$), $\frac{a_0}{2\pi c} \Delta\omega_{\text{VPC}}$ & $4.0 \times 10^{-2}$ & $2 |t_0|$ \\ [0.5ex]
 \hline
 splitting between the two spin-up armchair states, $\Delta_0$ & see Fig.~\ref{fig:parameters} & see Fig.~\ref{fig:parameters} \\ [0.5ex]
 \hline
\end{tabular}
\label{Tab:TB_parameters}
\end{table}

\begin{figure}[ht]
\centering
    \includegraphics[width=0.5\textwidth]{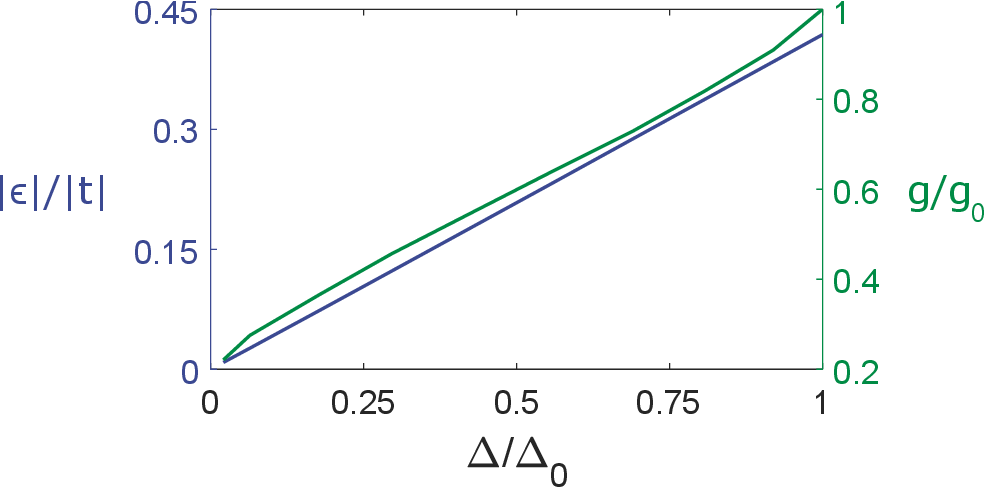}
    \caption{\label{fig:parameters}
    The relation between the tight-binding model parameter $\epsilon$, the on-site perturbation along the armchair domain wall  (left y axis, blue), and the photonic structural parameter $g$, the height of the gaps most adjacent to the armchair domain (right y axis, green), and the splitting between the two spin-up edge states $\Delta$.
    When $g = g_0$, the gaps are identical to those in the SPC$^{1,2}$ bulk, and $\Delta = \Delta_0 \approx 1.1 \times 10^{-2} (2\pi c/a_0)$.
    This splitting corresponds to $|\epsilon| \approx 0.42 |t|$ in the tight-binding model and $g = g_0 = 0.15a_0$ in the photonic design.
    }
\end{figure}

\section*{Microwave experiment setup}

\begin{figure}[ht]
    \centering
    \includegraphics[width=0.85\textwidth]{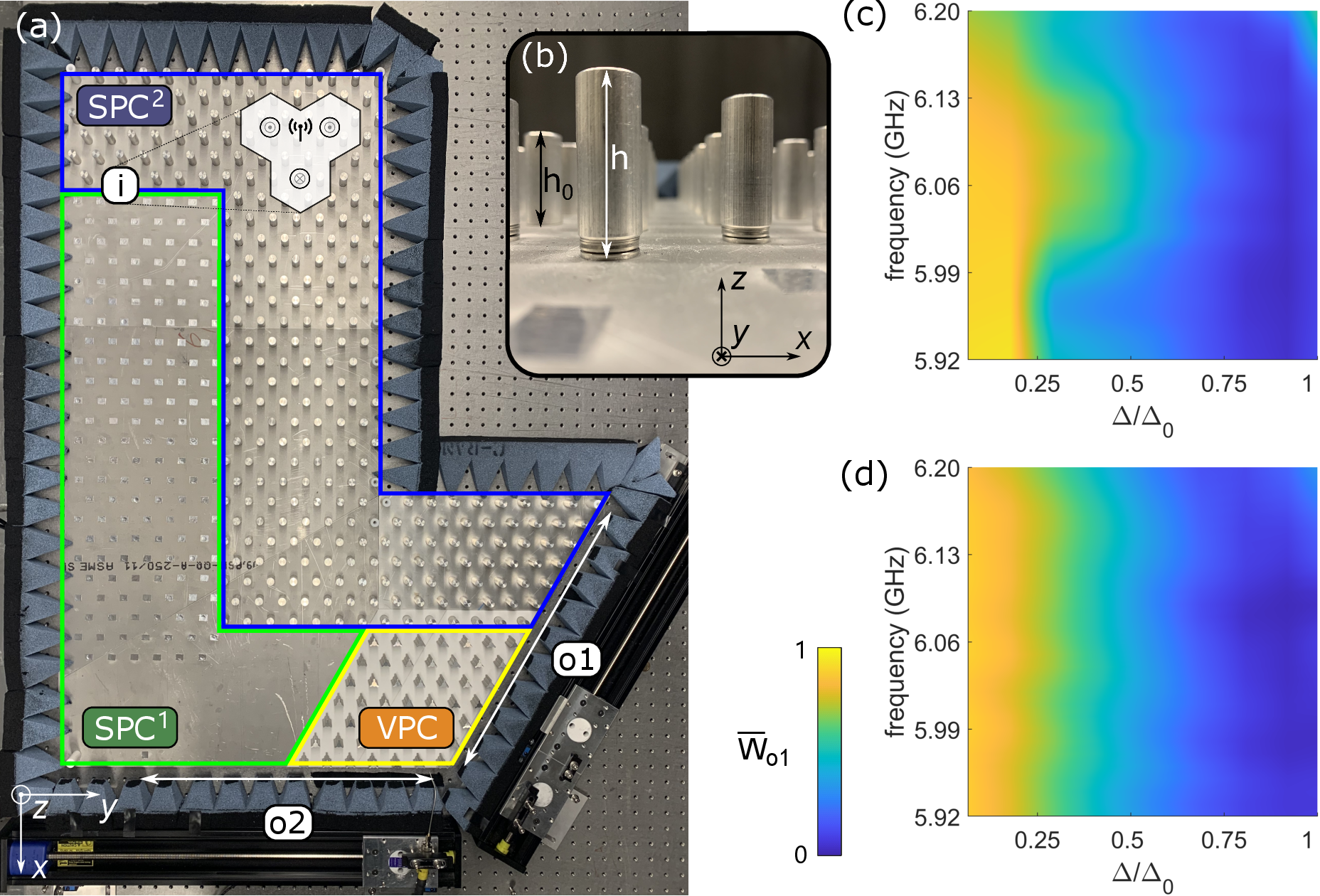}
    \caption{
    (a) The microwave experiment setup.
    We lift the top plate for visualization, so the aluminum rods in the SPC$^1$ region are not shown.
    The small squares in the SPC$^1$ region are pieces of aluminum foil tape used to seal holes on the bottom plate.
    The entire structure is surrounded by electromagnetic wave absorbers to eliminate the undesired reflection due to the external metallic structure.
    (b) Perturbed rods along the armchair domain wall.
    We install aluminum gaskets under the aluminum rods to increase $h$ and decrease $g$.
    (c,d) Experiment (c) and COMSOL simulation (d) results of the received energy ratio $\overline{W}_{o1} \equiv W_{o1} / \left( W_{o1}+W_{o2} \right)$ across the topological band gap for $9$ different $h$ configurations.
    }
    \label{fig:structure}
\end{figure}

Fig.~\ref{fig:structure} shows the structure used in the microwave experiment.
The dimensions of the SPC$^{1,2}$ and VPC unit cells are the same as in Ref.~\cite{SM_YLi:2022}.
The aluminum tripods of VPC are fabricated using wire electrical discharge machining.
The tripods between the two parallel plates are supported by two pieces of structural foam (ROHACELL 51HF).
We use another piece of structural foam with laser-cut holes to precisely locate the position of every tripod.

The source is a $23$mm long, $\hat{z}$-directional antenna made by trimming one lead of a coaxial cable.
The source is mounted at the position that ensures the $\Psi_{\mathbf{K} \uparrow}$ state is selectively excited (Fig.~\ref{fig:structure}a inset).
The excitation efficiencies of the two valley-polarizations satisfy $\eta_{\mathbf{K} \uparrow} / \eta_{\mathbf{K}' \uparrow} \approx 43$~\cite{SM_YLi:2022}.
(Note that the SPC$^{1,2}$ regions are flipped compared to those in Ref.~\cite{SM_YLi:2022}.)
Therefore, the valley $\mathbf{K}$ spin-up state is selectively excited with the setup in Fig.~\ref{fig:structure}.

We use the Keysight N5222A vector network analyzer to perform the measurement.
Each set of measurement contains the transmission spectra taken at $257$ uniformly spaced positions $r_i$ along the perophery at output o1 or o2 (Fig.~\ref{fig:structure}a).
The total amount of the received energy at a output port is calculated by summing $|S_{21}(r_i)|^2$ over all $r_i$ along that port.
The detect antenna is mounted on a motor-driven Velmex BiSlide rail.
The experiment and simulation results are shown in Fig.~\ref{fig:structure}c,d.
%The experiment result is averaged with a moving averaging window with width $\delta f=0.04$(GHz).

\section*{Analytical calculation}

An arbitrary state before entering the armchair domain wall can be written as $\Psi (l=0) = \cos(\theta/2) \Psi_{\mathbf{K}} + e^{i\varphi} \sin(\theta/2) \Psi_{\mathbf{K}'}$, where $l$ is the distance the state traversed along the armchair domain wall.
Here the spin-up labels are omitted for clearance.
Therefore, state exiting the armchair domain wall of length $L$ is
\begin{equation}\label{eq:state_evolution}
\Psi (l=L) = \left(\cos \frac{\theta}{2} + e^{i\varphi} \sin \frac{\theta}{2}\right) \frac{\Psi_+}{\sqrt{2}} e^{-i\omega_+ \frac{L}{v_g}} +
\left(\cos \frac{\theta}{2} - e^{i\varphi} \sin \frac{\theta}{2}\right) \frac{\Psi_-}{\sqrt{2}} e^{-i\omega_- \frac{L}{v_g}},
\end{equation}
where $\omega_{\pm} = \omega_0 \pm \Delta/2$.

Therefore, if the initial state is purely $\mathbf{K}$-polarized, e.g., $\theta=0$, the ratio of the $\mathbf{K}$($\mathbf{K}$')-polarization after exiting the armchair domain wall is,
\begin{equation}\label{eq:Rabi}
\begin{aligned}
& P_{\mathbf{K}} \equiv |\langle \Psi (l=L^{\text{TB}}_{\text{eff}}) | \Psi_{\mathbf{K}} \rangle|^2
= \frac{1}{2} + \frac{1}{2} \cos(L^{\text{TB}}_{\text{eff}} \Delta / v_g); \\
& P_{\mathbf{K}'} \equiv |\langle \Psi (l=L^{\text{TB}}_{\text{eff}}) | \Psi_{\mathbf{K}'} \rangle|^2
= \frac{1}{2} - \frac{1}{2} \cos(L^{\text{TB}}_{\text{eff}} \Delta / v_g).
\end{aligned}
\end{equation}
The effective armchair domain wall length $L^{\text{TB}}_{\text{eff}}$ for which the state $\Psi$ experiences the two-edge-state Hamiltonian is shorter than the actual armchair domain wall length, i.e., $L^{\text{TB}}_{\text{eff}} < L$,
because $\Psi$ does not turn an abrupt right angle when coupling-in from the zigzag domain wall to the armchair domain wall, or coupling-out from the armchair one to the zigzag one (see Fig.~3c-e in the main text).
$L^{\text{TB}}_{\text{eff}} \approx 21.2 a_0$ is obtained by fitting $P_{\mathbf{K}}$ to the tight-binding result (Fig.~3b in the main text).

For the photonic structure, first, because the excited state is not ideally $\mathbf{K}$-polarized, we use $\theta \approx 2\,\text{arccot} \sqrt{43}$ to describe the valley-selective source.
Moreover, because the photonic structure contains fewer unit cells than the tight-binding model, the finite-size effect in the VPC region becomes notable.
Consequently, when Port o1 receives $\Psi_{\mathbf{K}}$, it also receives a small portion of the decay tail of $\Psi_{\mathbf{K}'}$, and vice versa for Port o2~\cite{SM_YLi:2022}.
The energy received by Port o1 and o2 are
\begin{equation}\label{eq:energy_at_Ports}
\begin{aligned}
& W_{\text{o1}} = |\langle \Psi (l=L^{\text{sim}}_{\text{eff}}) | \Psi_{\mathbf{K}} \rangle|^2 W_{tr} + |\langle \Psi (l=L^{\text{sim}}_{\text{eff}}) | \Psi_{\mathbf{K}'} \rangle|^2 W_{tu}; \\
& W_{\text{o2}} = |\langle \Psi (l=L^{\text{sim}}_{\text{eff}}) | \Psi_{\mathbf{K}'} \rangle|^2 W_{tr} + |\langle \Psi (l=L^{\text{sim}}_{\text{eff}}) | \Psi_{\mathbf{K}} \rangle|^2 W_{tu},
\end{aligned}
\end{equation}
where $W_{tr}$ and $W_{tu}$ are the \textit{transmitted} and \textit{tunneled} energy that we discussed in Ref.~\cite{SM_YLi:2022}.
Similar to the tight-binding model, we obtain $L^{\text{sim}}_{\text{eff}}$ by fitting $\overline{W}_{o1} \equiv W_{\text{o1}} / \left( W_{\text{o1}}+W_{\text{o2}} \right)$ to the COMSOL simulation result.
This provides the semi-analytical calculation in Fig.~5b in the main text.

\section*{Design the perturbation in the photonic structure}

\begin{figure}[ht]
\centering
    \includegraphics[width=1\textwidth]{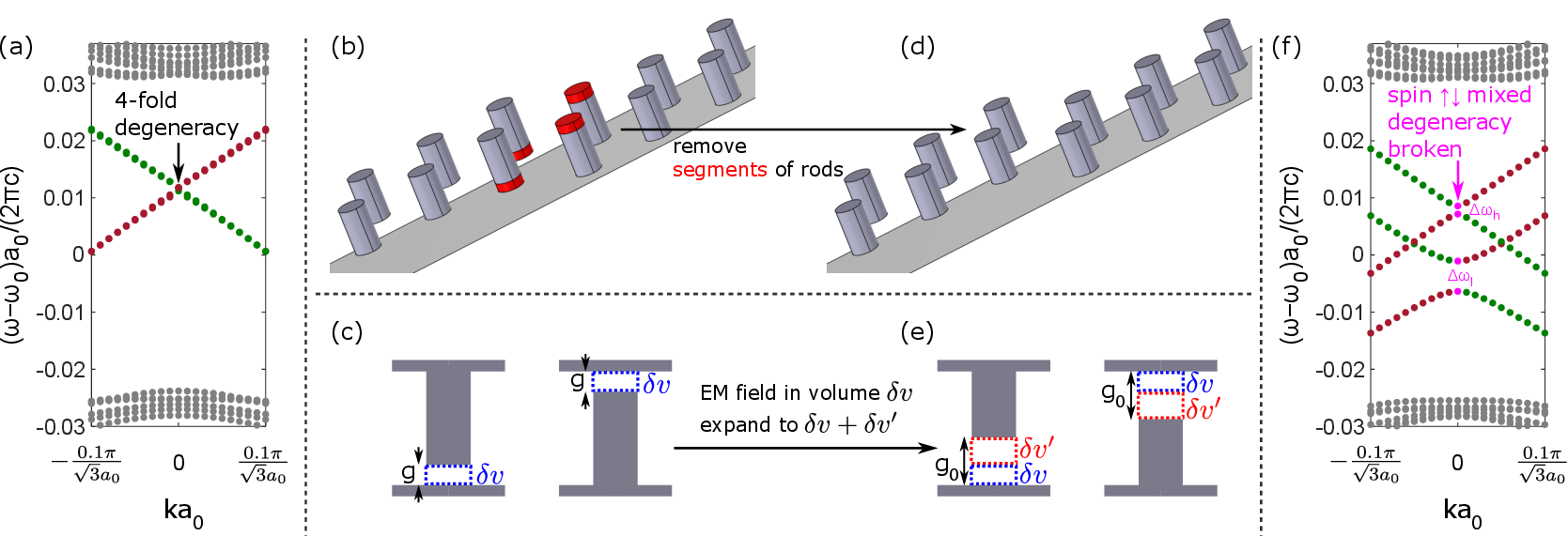}
    \caption{\label{fig:compare_armchair}
    Band diagrams and the structures of the armchair domain wall with $g = 0.22 g_0$ (a,b,c) and $g = g_0$ (d,e,f).
    The top PEC plate is hidden for better visualization.
    The schematic diagrams of the structure are not to scale.
    Increasing $g$ from $0.22g_0$ to $g_0$ is equivalent to removing the PEC in the shaded red region in (b).
    This causes the electromagnetic field within volume $\delta v$ (c, blue dashed line) to expand to volume $\delta v + \delta v'$ (e, blue and red dashed line).
    }
\end{figure}

We notice that when $g = 0.22g_0$, the armchair domain wall hosts a 4-fold degeneracy at the $\Gamma$ point (Fig.~\ref{fig:compare_armchair}a-c).
This configuration is the starting point for the perturbation-theory-based analysis.

The conventional cavity shape perturbation theory is used to analyze the degeneracy-breaking when a small volume, $\delta v$, of perfect electric conductor (PEC) is \textit{added} to the cavity~\cite{SM_Slater:1946,SM_Dombrowski:1984}:
The electromagnetic field within the volume $\delta v$ is removed, and the field elsewhere is assumed unchanged.
The system Hamiltonian after perturbation is $H = \omega_{p0} I - \Delta$, where $\omega_{p0}$ is the degenerate frequency before perturbation.
$\Delta$ is a matrix depending on the field overlap between different modes within the perturbation volume $\delta v$,
\begin{equation}
\Delta_{ij} = \iiint_{\delta v} \left( \epsilon_0 \mathbf{E}^*_i \cdot \mathbf{E}_j - \mu_0 \mathbf{H}^*_i \cdot \mathbf{H}_j \right) dv,
\end{equation}
where $i,j$ are mode indices.

In our structure, however, the degeneracy is broken by removing small pieces of PEC (Fig.~\ref{fig:compare_armchair}b,d).
Although the conventional cavity shape perturbation theory cannot be directly applied in this scenario, we extend its formalism to model the expansion of the eigenmode field distribution due to the removal of PEC segments.
We notice that most of the electromagnetic field is distributed in the gaps between the rod and the top (bottom) plate.
Removing segments of the PEC rods is equivalent to expanding those gap volumes from $\delta v$ to $\delta v + \delta v'$.
Therefore, the system Hamiltonian after perturbation becomes $H = \omega_{p0} I + \alpha \Delta$.
The minus sign is replaced by a plus sign because the field in the volume $\delta v$ is not removed but expanded.
$\alpha$ is a parameter describing the volume expansion.

While increasing the gap $g$ introduces inter-valley scattering, it also introduces inter-spin coupling in the proximity of the $\Gamma$ point.
The inter-spin coupling is manifested as two avoided crossings centered at $\left(\omega_h - \omega_0\right) = 0.009 (2\pi c)/a_0$ and $\left(\omega_l - \omega_0\right) = -0.007 (2\pi c)/a_0$, where the subscripts $h$ and $l$ stand for high- and low-frequency.

Here, we analyze the two avoided crossings using the ``extended" cavity perturbation theory.
When $g = 0.22 g_0$, both the inter-valley and the inter-spin couplings are negligible, and the armchair interface supports a 4-fold degeneracy (Fig.~\ref{fig:compare_armchair}a).
As the height of the rods most close to the interface decreases, the 4-fold degeneracy is broken (Fig.~\ref{fig:compare_armchair}f).

We evaluate the field overlap terms within the gap volume $\delta v$ (Fig.~\ref{fig:compare_armchair}c) and calculate matrix $\Delta$ in the spin-valley basis, $\left\{ \Psi_{\mathbf{K} \uparrow}, \Psi_{\mathbf{K}' \uparrow}, \Psi_{\mathbf{K} \downarrow}, \Psi_{\mathbf{K}' \downarrow} \right\}$.
The numerical result shows that all four diagonal entries in $\Delta$ are identical, and $\Delta$ is in the form of,
\begin{equation}
\overline{\Delta} = 
\begin{pmatrix}
 s & p & q & r \\
 p & s & r & q \\
 q & r & s & p \\
 r & q & p & s
\end{pmatrix} \approx
\begin{pmatrix}
 1.00  & -0.57 & -0.06 & 0.03 \\
 -0.57 & 1.00  & 0.03  & -0.06 \\
 -0.06 & 0.03  & 1.00  & -0.57 \\
 0.03  & -0.06 & -0.57 & 1.00
\end{pmatrix},
\end{equation}
where $\overline{\Delta} \equiv \Delta/diag(\Delta)$ and $diag(\Delta) \approx -2.9 \times 10^{-3} (2\pi c)/a_0$.
The volume expansion parameter $\alpha \approx 4$.

For convenience, we use four variables $s,p,q,r$ to represent the numerical value of $\Delta$.
Eigenfrequencies and eigenmodes of the system after removing segments of rods are in the following table.
\begin{table}[ht]
 \begin{tabular}{| c | c |} 
 \hline
 \qquad eigenfrequencies \qquad\qquad & \qquad\qquad\qquad eigenmodes \qquad\qquad\qquad\qquad \\ [0.5ex]
 \hline\hline
 $s+p+q+r$ & $\left( \Psi_{\mathbf{K} \uparrow} + \Psi_{\mathbf{K}' \uparrow}
+ \Psi_{\mathbf{K} \downarrow} + \Psi_{\mathbf{K}' \downarrow} \right) /2$ \\ [0.5ex]
 \hline
 $s+p-q-r$ & $\left( \Psi_{\mathbf{K} \uparrow} + \Psi_{\mathbf{K}' \uparrow}
- \Psi_{\mathbf{K} \downarrow} - \Psi_{\mathbf{K}' \downarrow} \right) /2$ \\ [0.5ex]
 \hline
 $s-p+q-r$ & $\left( \Psi_{\mathbf{K} \uparrow} - \Psi_{\mathbf{K}' \uparrow}
+ \Psi_{\mathbf{K} \downarrow} - \Psi_{\mathbf{K}' \downarrow} \right) /2$ \\ [0.5ex]
 \hline
 $s-p-q+r$ & $\left( \Psi_{\mathbf{K} \uparrow} - \Psi_{\mathbf{K}' \uparrow}
- \Psi_{\mathbf{K} \downarrow} + \Psi_{\mathbf{K}' \downarrow} \right) /2$ \\ [0.5ex]
 \hline
\end{tabular}
\label{Tab:eigenvalues_eigenvectors}
\end{table}

\begin{figure}[ht]
\centering
    \includegraphics[width=0.45\textwidth]{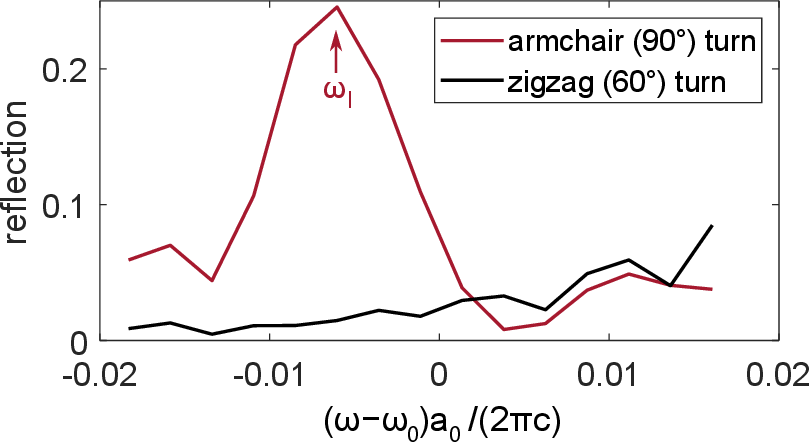}
    \caption{\label{fig:f_dependent_refl}
    Frequency-dependent reflection at an armchair ($90^{\circ}$) turn.
    A zigzag ($60^{\circ}$) turn is used as a baseline.
    The reflection increases at $\omega_l$, corresponding to the low-frequency avoided crossing due to inter-spin coupling.
    }
\end{figure}

The four eigenfrequencies correspond to the two avoided crossings at the $\Gamma$ point.
The higher avoided crossing is centered at $\omega_h = s+p$ with width $\Delta \omega_h = 2(q+r)$;
The lower avoided crossing is centered at $\omega_l = s-p$ with width $\Delta \omega_l = 2(q-r)$.
The two avoided crossings can hardly be suppressed to a negligible amount simultaneously because that requires $s = p = 0$.
With the perturbation of increasing the gap $g$, $q \approx -2r$ and $\Delta \omega_l \approx 3 \Delta \omega_h$ --  the higher avoided crossing is suppressed while the lower one is three times as wide, resulting in an increased back-reflection around $\omega_l$ (Fig.~\ref{fig:f_dependent_refl}).

\section*{Nonzero reflection and weak topological reflection}

The avoided crossings at $k=0$ and the resulting back-reflection indicate that the topological protection of the spin DoF is not strictly robust.
In general, the topological protection of photonic systems is weaker than their electronic counterparts:
In electronic QSH TIs, the Kramers’ degeneracy is protected by the time-reversal symmetry. In photonic QSH TIs, it is the crystalline symmetry and/or the TE/TM polarizations of the electromagnetic wave that help synthesize a pseudo-Kramer’s degeneracy.
Therefore, there is always an imperfection of the topological protection~\cite{SM_JHJiang:2019,SM_Agarwal:2023}.

\begin{figure}[ht]
\centering
    \includegraphics[width=0.8\textwidth]{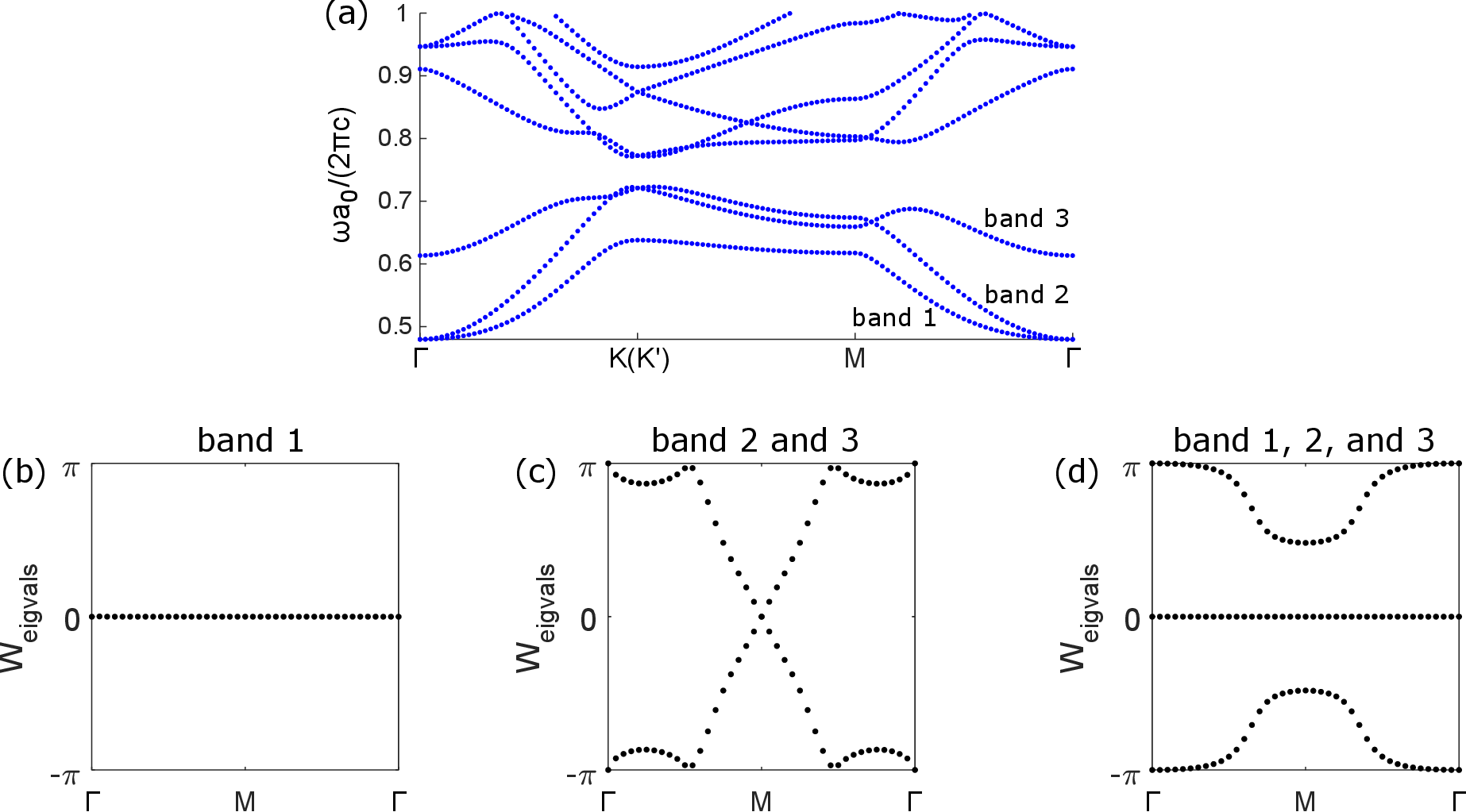}
    \caption{\label{fig:Wilson_loop_eigenvalues}
    (a) Photonic band diagram of the SPC$^{1,2}$.
    The three bands below the complete band gap are labeled as bands 1, 2, and 3 from low to high frequency.
    (b-d) The arguments (phases) of the Wilson loop eigenvalues of bands 1, 2, and 3 as functions of the momentum.
    (b) Only considering the isolated band 1, the Wilson loop eigenvalue has zero phase.
    (c) Only considering bands 2 and 3, the two Wilson loop eigenvalues wind non-trivially in opposite directions from $-\pi$ to $\pi$.
    (d) Considering all three bands, there is no signature of a non-trivial winding.
    }
\end{figure}

An analysis of the SPC$^{1,2}$ reveals the fragile topology~\cite{SM_Vishwanath:2018,SM_Bradlyn:2019} in their low-energy bands.
When only considering the two bands below the complete band gap, the arguments of the two Wilson loop eigenvalues wind in opposite directions from $-\pi$ to $\pi$ as functions of the momentum (Fig.~\ref{fig:Wilson_loop_eigenvalues}c).
However, when taking all three bands below the band gap into consideration, the nontrivial winding disappears (Fig.~\ref{fig:Wilson_loop_eigenvalues}d).

The nonzero reflection at the armchair channel at $k=0$ and the fragile topology suggest that this compact, low-reflection directional coupler does not rely on a strict topological protection.
In fact, all lossless waveguides, regardless of their topological classification, can be analyzed using the multi-modal perturbation theory method, and one can design specific interactions in a limited modal subspace by searching over all possible perturbations.
Nevertheless, the knowledge of topology and symmetry simplifies the design procedure:
We could identify the properties of all four modes based on their topological indices and directly conclude what interfaces cause robust transmission and what create the desired interaction without any computational design algorithms.
We also believe that when computational algorithms are deployed to design a photonic structure, the knowledge of the objective-required point group symmetries could accelerate the design procedure by drastically reducing the search space of all possible geometries~\cite{SM_Jelinek:2022}.

\section*{Realizing an arbitrary unitary operator}

A general operation in the SU($2$) group can be written as~\cite{SM_book_QCQI}
\begin{equation}\label{eq:SU2}
\begin{pmatrix}
e^{i(\alpha - \beta/2 - \delta/2)} \cos \frac{\gamma}{2} & -e^{i(\alpha - \beta/2 + \delta/2)} \sin \frac{\gamma}{2}\\
e^{i(\alpha + \beta/2 - \delta/2)} \sin \frac{\gamma}{2} & e^{i(\alpha + \beta/2 + \delta/2)} \cos \frac{\gamma}{2}
\end{pmatrix}.
\end{equation}

We demonstrate that this topological interferometer with additional topological waveguides can perform any arbitrary $2 \times 2$ unitary operation on the $\Psi_{\mathbf{K} \uparrow}$, $\Psi_{\mathbf{K}' \uparrow}$ states.

\begin{figure}[ht]
\centering
    \includegraphics[width=0.7\textwidth]{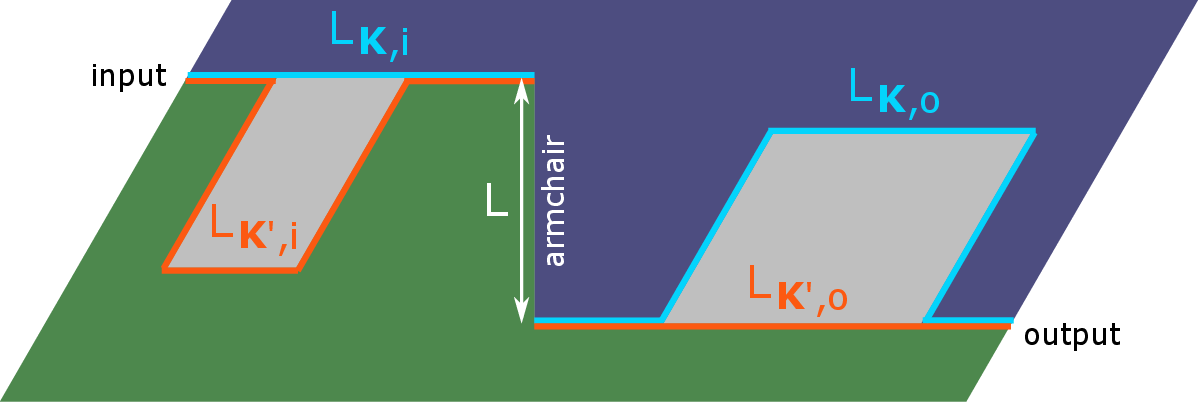}
    \caption{\label{fig:su2_system}
    Schematic of the structure realizing any arbitrary SU($2$) operation on the $\Psi_{\mathbf{K} \uparrow}$, $\Psi_{\mathbf{K}' \uparrow}$ states.
    VPC regions are add to separate the two spin-up state according to their valley indices.
    Hence, $\Psi_{\mathbf{K} \uparrow}$ and $\Psi_{\mathbf{K}' \uparrow}$ experience different optical paths both before and after interference ($L_{\mathbf{K},i} \neq L_{\mathbf{K}',i}$ and $L_{\mathbf{K},o} \neq L_{\mathbf{K}',o}$).
    }
\end{figure}

The evolution operator of the interferometer is
\begin{equation}\label{eq:t_evolution}
e^{-i H L/v_g} = e^{-i \omega_0 L/v_g}
\begin{pmatrix}
\cos L \Delta/(2 v_g) & -i \sin L \Delta/(2 v_g) \\
-i \sin L \Delta/(2 v_g) & \cos L \Delta/(2 v_g)
\end{pmatrix}.
\end{equation}

With the phase shift accumulated in the zigzag SPC$^+$-SPC$^-$ waveguides and VPC-SPC waveguides, the overall operation performed by the interferometer-waveguide system can be written as
\begin{equation}\label{eq:U_sys}
U_{\text{sys}} = e^{-i \omega_0 L/v_g}
\begin{pmatrix}
e^{i\omega_0 L_{\mathbf{K},o}/v_g} & 0 \\
0 & e^{i\omega L_{\mathbf{K}',o}/v_g}
\end{pmatrix}
\begin{pmatrix}
\cos L \Delta/(2 v_g) & -i \sin L \Delta/(2 v_g) \\
-i \sin L \Delta/(2 v_g) & \cos L \Delta/(2 v_g)
\end{pmatrix}
\begin{pmatrix}
e^{i\omega_0 L_{\mathbf{K},i}/v_g} & 0 \\
0 & e^{i\omega_0 L_{\mathbf{K}',i}/v_g}
\end{pmatrix},
\end{equation}
where $\omega_0 L_{\mathbf{K},o}/v_g$, $\omega_0 L_{\mathbf{K}',o}/v_g$ are the phases accumulated after exiting the interferometer;
$\omega_0 L_{\mathbf{K},i}/v_g$, $\omega_0 L_{\mathbf{K}',i}/v_g$ are those before entering the interferometer.

For convenience, we define
\begin{equation}
\phi_0 = \omega_0 \frac{L_{\mathbf{K},i} + L_{\mathbf{K}',o} + L}{v_g} - \frac{\pi}{2},\quad
\phi_1 = \omega_0 \frac{L_{\mathbf{K},o} - L_{\mathbf{K}',o}}{v_g} + \frac{\pi}{2},\quad
\phi_2 = \omega_0 \frac{L_{\mathbf{K}',i} + L_{\mathbf{K},i}}{v_g} + \frac{\pi}{2}.
\end{equation}
The extra $\pi/2$ terms are to compensate the imaginary unit $i$ in Eq.~\ref{eq:t_evolution}.

Eq.~\ref{eq:U_sys} consequently becomes
\begin{equation}\label{eq:U_sys_rewritten}
\begin{aligned}
U_{\text{sys}} & = i e^{i \phi_0}
\begin{pmatrix}
-i e^{i \phi_1} & 0 \\
0 & 1
\end{pmatrix}
\begin{pmatrix}
\cos L \Delta/(2 v_g) & -i \sin L \Delta/(2 v_g) \\
-i \sin L \Delta/(2 v_g) & \cos L \Delta/(2 v_g)
\end{pmatrix}
\begin{pmatrix}
1 & 0 \\
0 & -i e^{i \phi_2}
\end{pmatrix} \\
& = 
\begin{pmatrix}
e^{i(\phi_0 + \phi_1)} \cos L \Delta/(2 v_g) & -e^{i(\phi_0 + \phi_1 + \phi_2)} \sin L \Delta/(2 v_g) \\
e^{i\phi_0} \sin L \Delta/(2 v_g) & e^{i(\phi_0 + \phi_2)} \cos L \Delta/(2 v_g)
\end{pmatrix}
\end{aligned}
\end{equation}

We notice that Eq.~\ref{eq:U_sys_rewritten} can be directly re-written in the form of Eq.~\ref{eq:SU2} by letting
\begin{equation}
L \Delta / v_g = \gamma,\quad
\phi_0 = \alpha + \frac{\beta}{2} - \frac{\delta}{2},\quad
\phi_1 = -\beta,\quad
\phi_2 = \delta.
\end{equation}

\section*{Scaling up the TDCs and analyzing energy transmission efficiency}

\begin{figure}[ht]
\centering
    \includegraphics[width=0.8\textwidth]{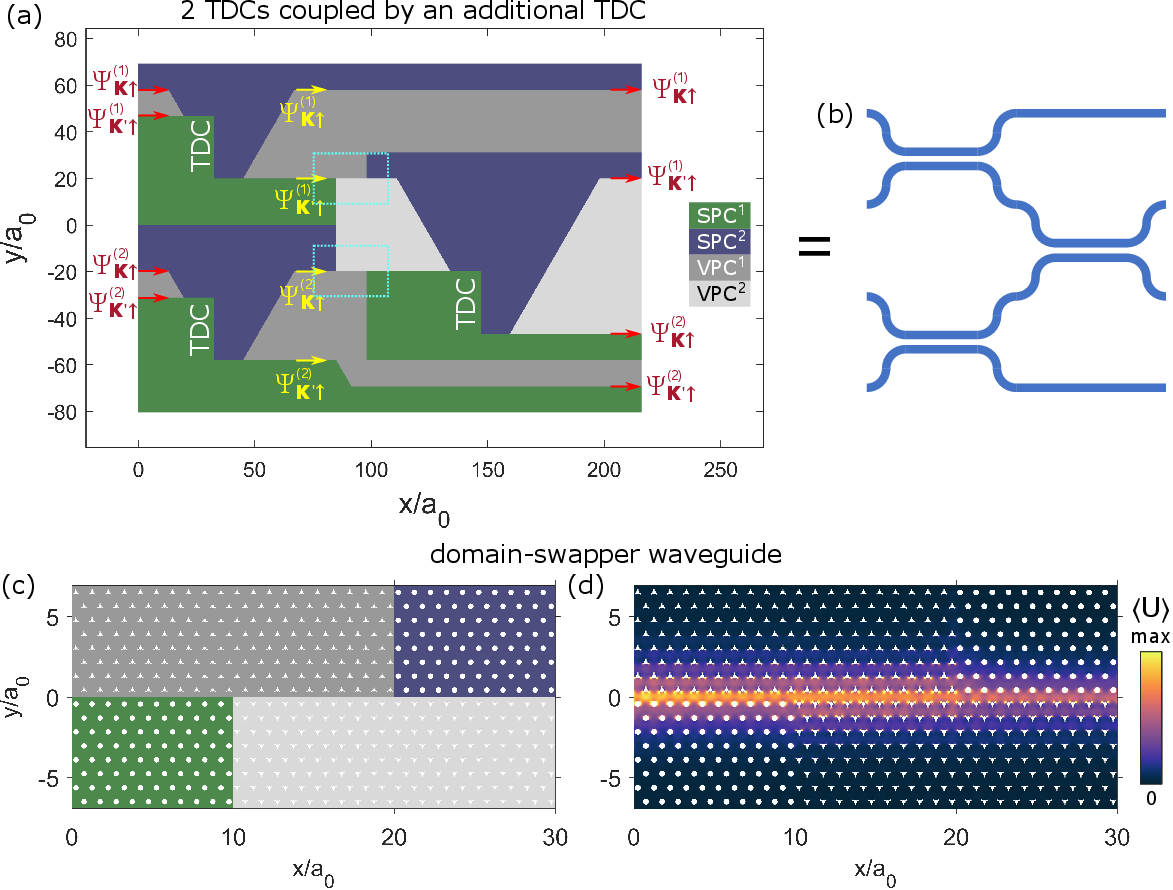}
    \caption{\label{fig:couple_two_TDCs}
    (a,b) Schematic diagram of two TDCs coupled by an additional TDC.
    The $\mathbf{K}(\mathbf{K}')$ valley states ($\Psi_{\mathbf{K}(\mathbf{K}')\uparrow}^{(\cdot)}$) at the first and the second TDC are labeled by the superscript.
    The blue dashed boxes enclose the domain-swapper waveguides.
    (c,d) The schematic (left) of and the transmitted photonic mode (right) along the single-mode domain-swapper waveguide.
    }
\end{figure}

To realize a $U(N)$ operation with a Mach–Zehnder interferometer (MZI) grid, properly coupled $N(N-1)/2$ directional couplers are necessary.
The first step is to couple two directional couplers with a third one.
Here, we provide a schematic of coupling two TDCs with an extra TDC (Fig.~\ref{fig:couple_two_TDCs}a).
To keep the coupling layout spatially compact, we introduce the design of single-mode “domain swapper” waveguides that use an additional domain of a different VPC to swap domains of different SPCs without altering the transmitted mode (Fig.~\ref{fig:couple_two_TDCs}c,d).

Following Clements et al.’s block implementation~\cite{SM_Clements:2016}, each input will travel through an $O(N)$ number of directional couplers, making the received energy reduced to $T^{O(N)}$ of the input, where $T$ is the transmission of one directional coupler.
Here, we investigate the transmission of one TDC.

\begin{figure}[ht]
\centering
    \includegraphics[width=0.9\textwidth]{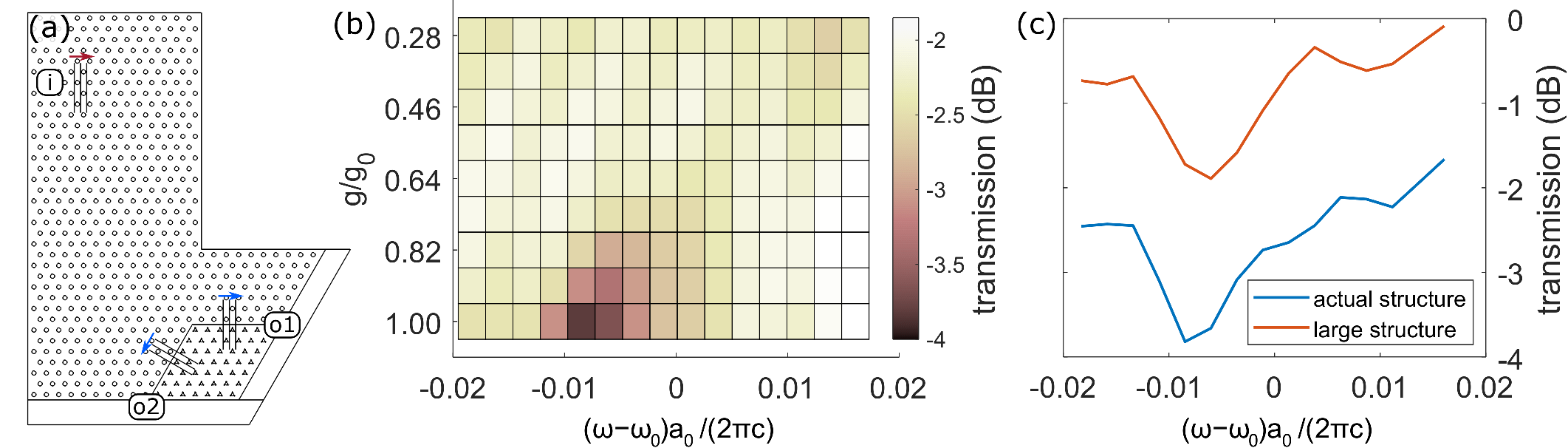}
    \caption{\label{fig:f_l_dependent_transmission}
    (a) Schematic of the TDC and the transverse planes for measuring the power flow.
    (b) Frequency-dependent transmission for nine different configurations.
    (c) Transmission of the actual structure (the insulating bulk is 7 unit cells deep along the transverse direction, as shown in (a)) and a large structure (the insulating bulk is 14 unit cells deep) when $g/g_0=0.91$.
    }
\end{figure}

We calculate the perpendicular time-averaged Poynting vector integrated over the transverse intersecting planes (corresponding to output ports o1,o2, and the input port i),
$\langle P_{\hat{\mathbf{n}}}^{port}\rangle \equiv \int_{port} da \langle \mathbf{S} \rangle \cdot \hat{\mathbf{n}}$ (for each port, the result is averaged over three adjacent planes, Fig.~\ref{fig:f_l_dependent_transmission}a),
and evaluate the transmission as $T=\left( \langle P_{\hat{\mathbf{n}}}^{o1} \rangle +\langle P_{\hat{\mathbf{n}}}^{o2} \rangle\right)/\langle P_{\hat{\mathbf{n}}}^{i} \rangle$.
The frequency-dependent result is presented in Fig.~\ref{fig:f_l_dependent_transmission}b.
At high frequencies, $\left(\omega-\omega_0\right)a_0/\left(2\pi c\right) \geq 0.002$ (higher than the lower avoided crossing), the transmission is consistently higher than $-2.2$dB.

Furthermore, we notice that a majority of the loss is due to the finite-size effect of the structure --- the insulating bulk is not large enough to shield all the loss through the exponential tail along the transverse direction.
After increasing the depth of the insulating bulk from 7 to 14, we find that the transmission significantly increased.
Fig.~\ref{fig:f_l_dependent_transmission}c shows that for the most lossy case ($g/g_0=0.91$, when the lower avoid crossing is the widest), doubling the insulating depth increases the transmission by $\sim 1.6$dB.

\begin{figure}[ht]
\centering
    \includegraphics[width=1.0\textwidth]{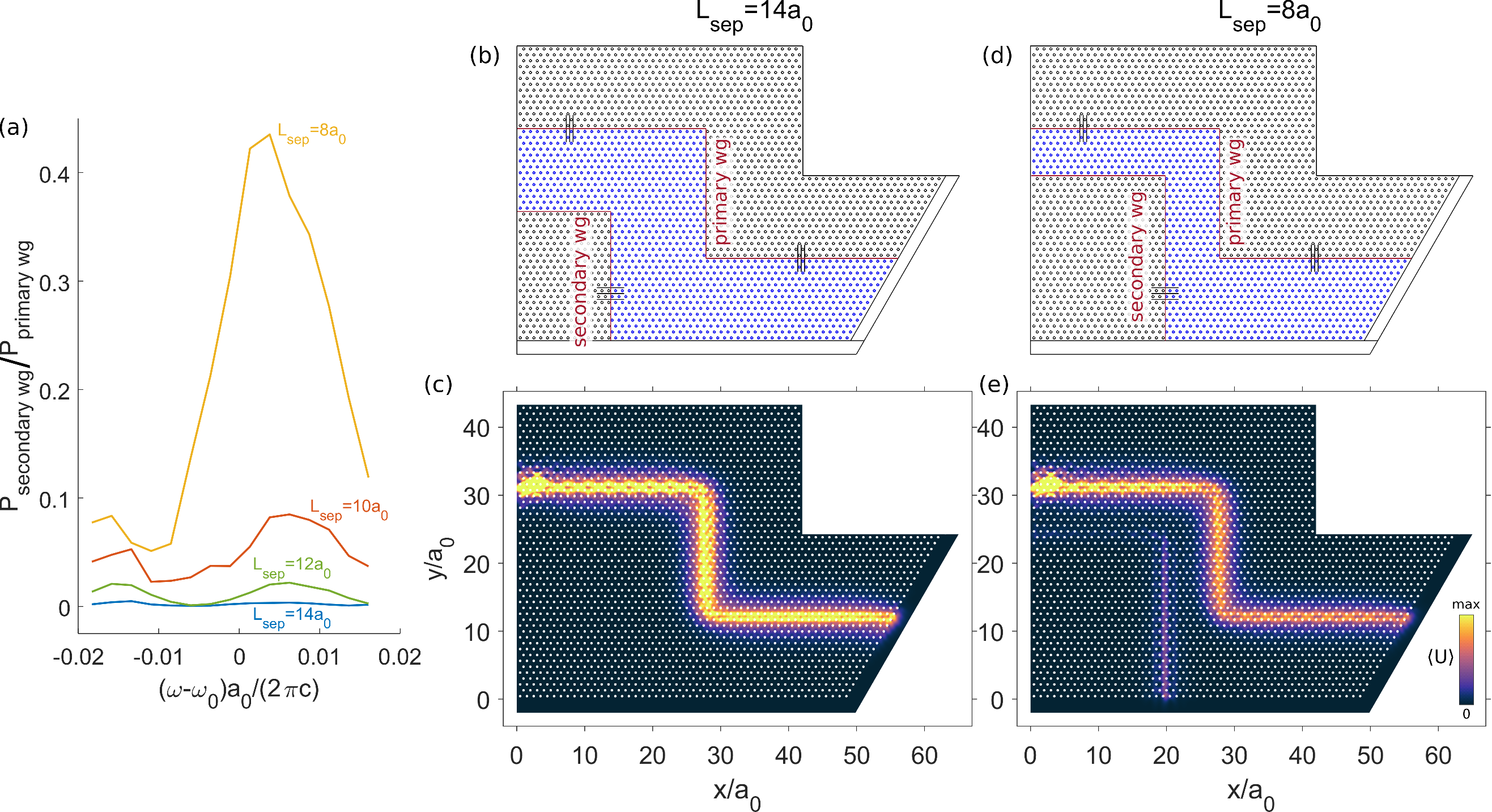}
    \caption{\label{fig:measure_cross_talk}
    Crosstalk between the primary and secondary waveguides.
    (a) The energy flow ratio as a function of the waveguide separation distance.
    (b,c) The two waveguides are separated by $L_{\text{sep}} = 14a_0$.
    The coupling from the primary waveguide to the secondary one is negligible.
    (d,e) $L_{\text{sep}} = 8a_0$.
    About $30\%$ energy in the primary waveguide couples to the secondary one.
    The time-averaged energy distributions in (c,e) are at $\omega = \omega_0$, where $\omega_0 \approx 0.75 (2\pi c/a_0)$.
    }
\end{figure}

When scaling up multiple TDCs, one also needs to consider how close two TDCs can be placed in parallel without compromising the functionality.
When a secondary waveguide is too close to the TDC, optical energy can couple to that secondary waveguide through the evanescent tail.
In an SPC waveguide, the propagating mode decays as $e^{-\kappa r_{\perp}}$ along the transverse direction, where $1/\kappa \approx 2.66a_0$~\cite{SM_YLi:2022}.
Although this decay tail is short, the crosstalk between two parallel waveguides could still be significant if the propagation length is sufficiently long (e.g., $20$--$30a_0$ for TDCs).
To examine the problem, we vary the separation between the primary and a secondary TDC from $8a_0$ to $14a_0$ and measure the energy in the two waveguides after light traverses through the primary TDC.
Our finding shows that crosstalk is negligible when the separation exceeds $12a_0$ (Fig.~\ref{fig:measure_cross_talk}).

\section*{Footprint of directional couplers}

\begin{figure}[ht]
\centering
    \includegraphics[width=0.7\textwidth]{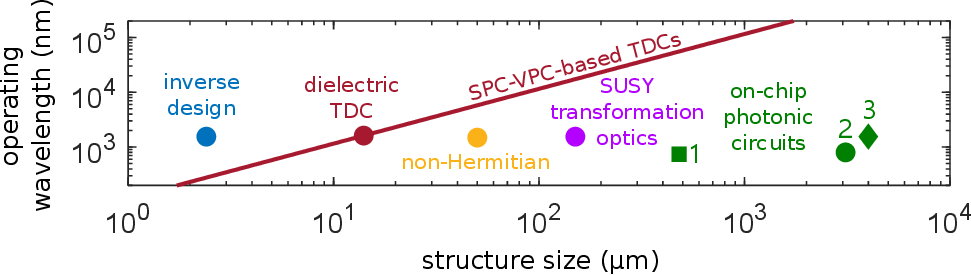}
    \caption{\label{fig:compare_structure_size}
    Comparing the sizes and operating wavelengths of different designs of directional couplers, beam splitters, and routing structures.
    Red line: the SPC-VPC-based TDCs with different lattice constants.
    Red dot: the TDC based on the dielectric SPCs introduced in the supplementary material of Ref.~\cite{SM_Kang:2018}.
    Green dots: designs based on on-chip photonic circuits. 1, 2, and 3 refer to Refs.~\cite{SM_Dong:2022,SM_OBrien:2008,SM_KHLuo:2019}, respectively.
    Magenta dot: the SUSY transformation optical mode converter in Ref.~\cite{SM_Yim:2022}.
    Yellow dot: the non-Hermitian steering structure in Ref.~\cite{SM_LiangFeng:2019}.
    Blue dot: the inverse-designed polarization splitter in Ref.~\cite{SM_Menon:2015}.
    }
\end{figure}

With the all-dielectric SPCs introduced in the supplementary information of Ref.~\cite{SM_Kang:2018}, the SPC-VPC-based TDCs could potentially be realized in the visible or telecom regime.
Fig.~\ref{fig:compare_structure_size} compares the footprints of TDCs, conventional, and newly emerging directional couplers/beam splitters.
TDCs are more compact than conventional directional couplers~\cite{SM_Dong:2022,SM_OBrien:2008,SM_KHLuo:2019}, which use transition regions with large radii of curvature to suppress back-reflection.
Two unconventional mechanisms, supersymmetric (SUSY) transformation optics and non-Hermitian optics show promise for innovative coupler designs. 
Compared to SUSY transformation optical mode-converters~\cite{SM_Walasik:2019,SM_Yim:2022}, TDCs are about one order of magnitude shorter because the former requires an adiabatic modification of the refractive index along the propagation direction;
Compared to non-Hermitian steering/routing designs~\cite{SM_LiangFeng:2019}, TDCs are only marginally smaller.
Novel computational design methods, e.g., inverse design, archive the smallest functioning beam-splitting structure~\cite{SM_Menon:2015}.

\bibliographystyle{unsrt}

\end{document}